\documentclass[amsmath,amssymb, aps, prb, twocolumn, notitlepage]{revtex4-2}
\pdfoutput=1
\usepackage[utf8]{inputenc}
\usepackage[export]{adjustbox}
\usepackage{multirow}
\usepackage{braket}
\usepackage{indentfirst}
\usepackage{graphicx}
\usepackage{dcolumn}
\usepackage{bm}
\usepackage[usenames]{color}
\usepackage{slashed}
\usepackage[utf8]{inputenc}
\usepackage{amsfonts}
\usepackage{amsmath}
\usepackage{amssymb}
\usepackage{hyperref}
\usepackage{latexsym}
\usepackage{color}
\usepackage{soul}
\usepackage{appendix}

\begin{document}
\title{Nontrivial worldline winding in non-Hermitian quantum systems}
\author{Shi-Xin Hu}
\affiliation{International Center for Quantum Materials, School of Physics, Peking University, Beijing, 100871, China}
\author{Yongxu Fu}
\email{yongxufu@pku.edu.cn}
\affiliation{International Center for Quantum Materials, School of Physics, Peking University, Beijing, 100871, China}
\author{Yi Zhang}
\email{frankzhangyi@gmail.com}
\affiliation{International Center for Quantum Materials, School of Physics, Peking University, Beijing, 100871, China}

\date{\today}

\begin{abstract}
Amid the growing interest in non-Hermitian quantum systems, non-interacting models have received the most attention. Here, through the stochastic series expansion quantum Monte Carlo method, we investigate non-Hermitian physics in interacting quantum systems, e.g., various non-Hermitian quantum spin chains. While calculations yield consistent numerical results under open boundary conditions, non-Hermitian quantum systems under periodic boundary conditions observe an unusual concentration of imaginary-time worldlines over nontrivial winding and require enhanced ergodicity between winding-number sectors for proper convergences. Such nontrivial worldline winding is an emergent physical phenomenon that also exists in other non-Hermitian models and analytical approaches. Alongside the non-Hermitian skin effect and the point-gap spectroscopy, it largely extends the identification and analysis of non-Hermitian topological phenomena to quantum systems with interactions, finite temperatures, biorthogonal basis, and periodic boundary conditions in a novel and controlled fashion. Finally, we study the direct physical implications of such nontrivial worldline winding, which bring additional, potentially quasi-long-range contributions to the entanglement entropy. 
\end{abstract}

\maketitle

\section{Introduction}

Recent explorations of non-Hermitian quantum systems have broadened the scope of condensed matter physics \cite{ashida2020,bergholtzrev2021,PhysRevLett.116.133903,leykam2017edge,torres2018anomalous,shen2018,kunst2018,yao2018,yao201802,gong2018,kawabataprx,lee2019an,yokomizo2019,longhi2019,origin2020,slager2020,yang2020,zhang2020,kawabata2021,xue2021simple,edgeburst2022,longhi2022,longhihealing,fu2023ana}, and rapidly spread to the field of higher-order non-Hermitian systems \cite{kawabata2019second,lee2019ho,edvardsson2019,kawabatahigher,okugawa2020,fu2021,yu2021ho,palacios2021,st2022} and exceptional points \cite{kawabata2019,yokomizo2020,jones2020,zhang2020ep,xue2020dirac,yang2021,denner2021,fu2022,mandal2021ep,delplace2021ep,liu2021ep,marcus2021ep,ghorashi2021dirac,ghorashi2021weyl}. Originating from effective models for open systems \cite{open1, open2, open3, open4, open5, open6, open7, open8, open9}, dissipative optical systems \cite{optical1, optical2, optical3, optical4, optical5, optical6}, and electric circuits \cite{circuit1, circuit2, circuit3, circuit4, circuit5, circuit7}, etc., non-Hermitian quantum systems display a wide range of interesting physical properties open to theoretical studies and experimental realizations. For example, the non-Hermitian skin effect (NHSE) is a remarkable feature that predicts an extensive number of eigenstates localized at the edges under open boundary conditions (OBCs) as well as the breakdown of the Bloch band theory \cite{PhysRevLett.116.133903, yao2018, open6, yokomizo2019, zhang2020, yang2020, origin2020}. 

Interestingly, the NHSE is also deeply associated with the nontrivial point-gap topology of non-Hermitian quantum systems, i.e., the winding number of the energy spectra under PBCs around the reference energy in the complex plane controls the occurrence or absence of the NHSE \cite{zhang2020, origin2020} and reflects non-Hermitian bulk-boundary correspondence \cite{yang2020,zirnstein2021,zirnstein02021}. Simultaneously, the NHSE must accompany the departure of the energy spectra under OBCs and PBCs \cite{zhang2020, origin2020}. However, the NHSE also comes with its systematic limitations: It focuses on the right-eigenstates of non-interacting fermion systems under OBCs, thus inapplicable to finite temperatures, interactions, periodic boundary conditions, and expectation values under biorthogonal bases, which are common scenarios in condensed matter physics. 

Beyond single-particle physics, researches on non-Hermitian quantum systems with interactions have also been picking up paces lately and revealed many exotic many-body properties \cite{PhysRevB.101.121109, manybody1, manybody2, manybody3, manybody4, manybody5, manybody6, manybody7, manybody8, manybody9, manybody10, manybody11, manybody12, manybody13, manybody14, manybody15, manybody16, manybody17}. Here, we take a quantum many-body perspective into non-Hermitian physics by generalizing the stochastic series expansion quantum Monte Carlo (SSE-QMC) \cite{PhysRevB.43.5950, AWSandvik_1992, sandvik2019stochastic, sandvik1999operatorloop} method to certain non-Hermitian quantum systems without the sign problem. The SSE-QMC method stochastically samples imaginary-time operator sequences, i.e., worldlines in $(D+1)$-dimensional space-time, in the Taylor series expansion of the partition function; it is highly efficient and easily implementable for some quantum spin \cite{SSEspin1, SSEspin2,SSEspin3,SSEspin4,SSEspin5,SSESpinandBoson} and boson lattice models \cite{SSESpinandBoson, SSEboson1,SSEboson2,SSEboson3}, albeit Hermitian or not. We obtain consistent results on non-Hermitian quantum many-body systems under OBCs. Under periodic boundary conditions (PBCs), however, the worldlines are dominated by nontrivial winding-number sectors and may obstruct convergence. To enhance ergodicity and facilitate convergence, we introduce a simple remedy for the SSE-QMC algorithm.

Importantly, like the NHSE, the nontrivial worldline winding may act as a defining character for non-Hermitian point-gap topological phenomena. In non-interacting cases, the nontrivial worldline winding corresponds to a nonzero point-gap topological invariant around the reference point $E_P=0$. However, unlike the NHSE \cite{PhysRevB.101.121109}, nontrivial worldline winding is also applicable for interacting quantum systems and finite temperatures; indeed, its emergence exhibits explicit interaction dependence. Also, the related phenomena are reflected in physical observables corresponding to biorthogonal expectation values, including additional contributions to the entanglement entropy that resembles quasi-long-range entanglement. Further, instead of a binary ``yes or no" answer, it offers a semi-quantitative measure of the extent of non-Hermitian topological physics at play. Finally, its PBC promptly complements the NHSE under OBCs. 

We organize the rest of this paper as follows: In the next section, we briefly review the SSE-QMC technique (Sec. \ref{sec:IIA}) and the non-Hermitian quantum physics (Sec. \ref{sec:IIB}) before examining SSE-QMC generalization and applicability on non-Hermitian quantum systems (Sec. \ref{sec:IIC}); then, in Sec.  \ref{sec:IID}, we discuss the results of non-Hermitian quantum spin chains under OBCs as examples. In Sec. \ref{sec:IIIA}, we show the difficulties that SSE-QMC calculations encounter for the same non-Hermitian quantum spin chains yet under PBCs; for the explanation, we discuss the nontrivial worldline winding in a non-Hermitian toy model in Sec. \ref{sec:IIIB}; correspondingly, we propose a simple algorithmic technique to enhance ergodicity in Sec. \ref{sec:IIIC}, which indeed restores the SSE-QMC credibility for non-Hermitian quantum models under PBCs. In Sec. \ref{sec:IIID}, we give a systematic analysis of the nontrivial worldline winding, whose conditions are consistent with the point-gap topology, as well as finite-temperature and interacting scenarios beyond the previous theoretical framework. Sec. \ref{sec:IV} is attributed to physical implications of such nontrivial worldline winding - additional contributions to the entanglement entropy. We summarize and conclude the paper in Sec. \ref{sec:V}, discussing potential generalizations such as general algorithms, higher dimensions, diverse boundary conditions, and other non-Hermitian topology.

\section{SSE-QMC method for non-Hermitian quantum systems}\label{sec:II}

\subsection{Review of the SSE-QMC method}\label{sec:IIA}

The SSE-QMC method is a powerful tool for calculating the physical quantities of quantum many-body systems. It is based upon the Taylor expansion of the Boltzmann factor in the partition function \cite{AWSandvik_1992}:
    \begin{equation}
Z=\operatorname{Tr}\left\{e^{-\beta \hat H}\right\}=\sum_\alpha \sum_{n=0}^{\infty} \frac{\beta^n}{n !}\left\langle\alpha\left|(-\hat H)^n\right| \alpha\right\rangle, \label{HermitianPartition}
\end{equation}
where $\beta$ is the inverse temperature, and $\left\{|\alpha\rangle\right\}$ is an orthogonal basis, e.g., $|\alpha\rangle=\left|S_1^z, S_2^z, \ldots, S_N^z\right\rangle$ for a spin system with $N$ sites. 

We can decompose the Hamiltonian $\hat H$ into:
\begin{equation}
\hat H=-\sum_{a, b} \hat H_{a,b},
\end{equation}
where $b$ labels different bonds (sites) within the lattice, and $a$ denotes different types of operators. Consequently, we re-express the partition function as:
\begin{equation}
Z=\sum_{\alpha} \sum_{n=0}^{\infty} \sum_{S_n} \frac{\beta^n}{n !}\langle\alpha| \prod_{i=1}^n \hat H_{a_i, b_i}|\alpha\rangle,
\end{equation}
where $\sum_{S_n}$ sums over different sequences of operators:
\begin{equation}
S_n=\left[a_1, b_1\right],\left[a_2, b_2\right], \ldots,\left[a_n, b_n\right].
\end{equation}

In practice, we truncate the Taylor series at a sufficiently large $M$ so that $M>n$ for the highest power with meaningful contribution, achieved via thermalization before the actual sampling. Instead of varying $n$, it is more convenient to consider an operator sequence with a fixed length $M$, including $n$ nontrivial operators and $M-n$ identity operators $\hat H_{0,0}=\hat I$ \cite{PhysRevB.43.5950}. Although the identity operators make no direct contribution, there are $M!/(M-n)!n!$ number of ways of equivalent insertions of such identity operators, a binomial factor we must divide out for the partition function:
\begin{equation}
Z=\sum_\alpha \sum_{S_M} \frac{\beta^n(M-n) !}{M !}\left\langle\alpha\left|\prod_{i=1}^M \hat H_{a_i, b_i}\right| \alpha\right\rangle, \label{eq:sseZ}
\end{equation}
where the operator sequence $S_M$ includes $n$ nontrivial operators and $M-n$ identity operators. 

\begin{figure}
    \centering
    \includegraphics[width=1 \linewidth]{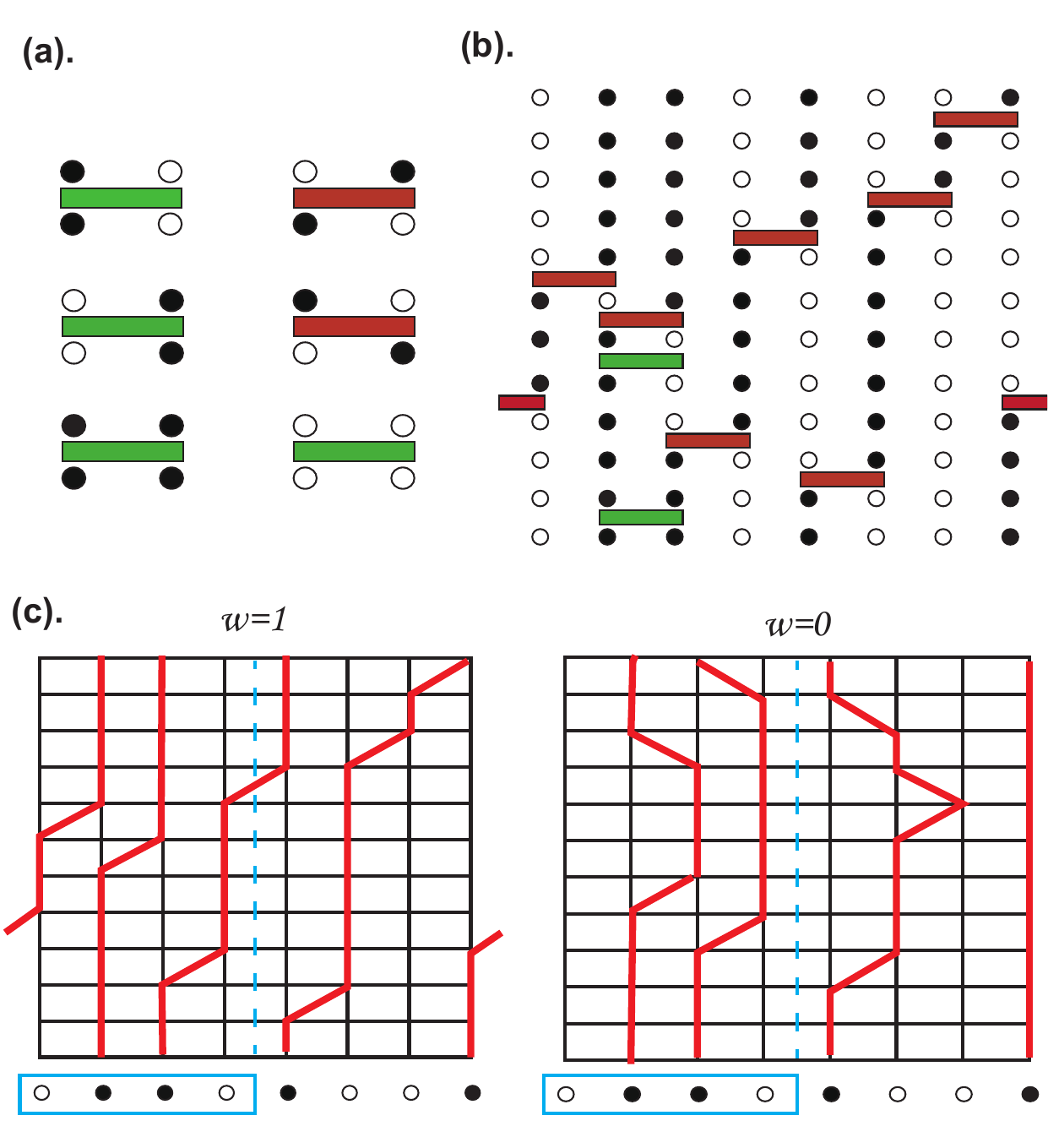}
    \caption{We illustrate key elements of the SSE-QMC method with a quantum spin model under PBC. (a) The vertices are finite matrix elements of operators; diagonal operators (green rectangles) keep the configurations intact from their lower legs at time slice $p-1$ to upper legs at $p$, while off-diagonal operators (red rectangles) alter the configurations, e.g., a spin swap. (b) We can obtain the configuration $\{|\alpha_p\rangle\}$ from the initial state $|\alpha\rangle$ and the operator sequence $S_M$; the imaginary time $p$ is vertical, and the lattice position $b$ is horizontal. The black (white) circle denotes a spin-up (down) site. (c) The worldlines following spin-up sites form closed loops that wrap around the system. The worldlines on the left corresponding to (b) possess a  winding number $w=1$ while the ones on the right show $w=0$. Nontrivial winding guarantees worldlines crossing boundaries (vertical dashed line) and influences the quantum entanglement between subsystems, such as $A$ (blue box) and $\Bar{A}$, as discussed in Sec. \ref{sec:IV}.}
    \label{vertexandConfig}
\end{figure}

It is convenient to define a propagated state \cite{PhysRevB.43.5950}:
\begin{equation}
|\alpha_p\rangle \propto \prod_{i=1}^p \hat H_{a_i, b_i}|\alpha\rangle,
\end{equation}
which satisfies the no-branching condition, i.e., $|\alpha_p\rangle$ is always proportional to one of the states in the chosen basis. Depending on whether the operator $\hat H_{a_p,b_p}$ is diagonal or off-diagonal, $|\alpha_p\rangle = \hat H_{a_p,b_p} |\alpha_{p-1}\rangle$ may either equal $|\alpha_{p-1}\rangle$ or differ from $|\alpha_{p-1}\rangle$ on the $b_{p}$ bond (site), e.g., due to spin flips. The identity (operator) is also diagonal. The finite matrix elements $\langle\alpha_p|\hat H_{a_p,b_p}|\alpha_{p-1}\rangle$ of the operators, also called the vertices and illustrated in Fig. \ref{vertexandConfig}(a), keep track of the configuration differences, if any, between two neighboring time slices $p-1$ and $p$; see examples in Fig. \ref{vertexandConfig}(a). 

We may sample the $|\alpha_p\rangle$ configurations in the $(D+1)$-dimensional space-time, uniquely determined by the initial state $|\alpha\rangle$ and the operator sequence $S_M$, with which we can trace $|\alpha_p\rangle$ along the imaginary-time direction slice-by-slice; see Fig. \ref{vertexandConfig}(b). Following Eq. \ref{eq:sseZ}, the Monte Carlo weight of each configuration is:
\begin{eqnarray}
    W(\alpha,S_M)&=&\frac{\beta^n(M-n)!}{M!}\left\langle\alpha\left|\prod_{p=1}^M \hat H_{a_p, b_p}\right| \alpha\right\rangle \nonumber \\
    &=&\frac{\beta^n(M-n) !}{M !} \prod_{p=1}^M \left\langle\alpha_{p}\left|\hat H_{a_p, b_p}\right| \alpha_{p-1}\right\rangle, \label{eq:sampleprob}
\end{eqnarray}
where $|\alpha_0\rangle = |\alpha_M\rangle = |\alpha\rangle$. As a result, we can evaluate the expectation value of operator $\hat{A}$ as:
\begin{equation}
\langle\hat{A}\rangle=\frac{\sum_{\alpha, S_M} A\left(\alpha, S_M\right) W\left(\alpha, S_M\right)}{\sum_{\alpha, S_M} W\left(\alpha, S_M\right)}, \label{MCsum}
\end{equation}
where $A\left(\alpha, S_M\right)$ is the matrix element of $\hat{A}$ given the configuration in $|\alpha\rangle$ and $S_M$. One important example is the expectation value $\langle \hat H_{a,b}\rangle$, where $H_{a,b}(\alpha,S_M)=n_{a,b}/\beta$, and $n_{a,b}$ is the number of $\hat H_{a,b}$ in the operator sequence $S_M$. 

There is one more essential requirement to make the SSE-QMC method work: the sampling probabilities $W(\alpha,S_M)$ in Eq. \ref{eq:sampleprob} [$W(\alpha,S_M)/\sum_{\alpha, S_M} W\left(\alpha, S_M\right)$ after normalization] need to be positive-semidefinite. Correspondingly, either the matrix elements $\langle\alpha_p|\hat H_{a_p,b_p}|\alpha_{p-1}\rangle$ are positive-semidefinite, or the number of negative matrix elements in the operator sequence is always even, so the overall product is still positive-semidefinite \cite{PhysRevB.43.5950}. If the negative probability cannot be removed by any means, we encounter the sign problem \cite{pan2022sign} and cannot carry out the calculations in a controlled way, especially for large systems. 

As the imaginary time propagates and we keep track of the configuration changes, e.g., the spin-up positions intervened by off-diagonal operators in a quantum spin model, we obtain a series of trajectories called the worldlines; see Fig. \ref{vertexandConfig}(c). The worldlines offer another representation of the configurations and play a crucial role in efficient loop updates for the SSE-QMC method \cite{PhysRevB.57.13382}. 

Due to the presenting trace in the partition function, $|\alpha_0\rangle = |\alpha_M\rangle$, the worldlines in the SSE-QMC samples must obey periodic boundary conditions in the imaginary-time direction and form closed loops [Fig. \ref{vertexandConfig}(c)]. Meanwhile, the worldlines can wrap around the system, and the net number of times they wrap around is called the winding number $w$ \cite{PhysRevB.57.13382}. $w$ can be a finite integer in PBCs, while $w$ should always be zero in OBCs. One of this work's key conclusions is the emergence of nontrivial dominant worldline winding in non-Hermitian quantum systems.

\subsection{Review of non-Hermitian physics: model, skin effect, and gap topology}\label{sec:IIB}

\begin{figure*}
    \centering
    \includegraphics[width = 1\linewidth]{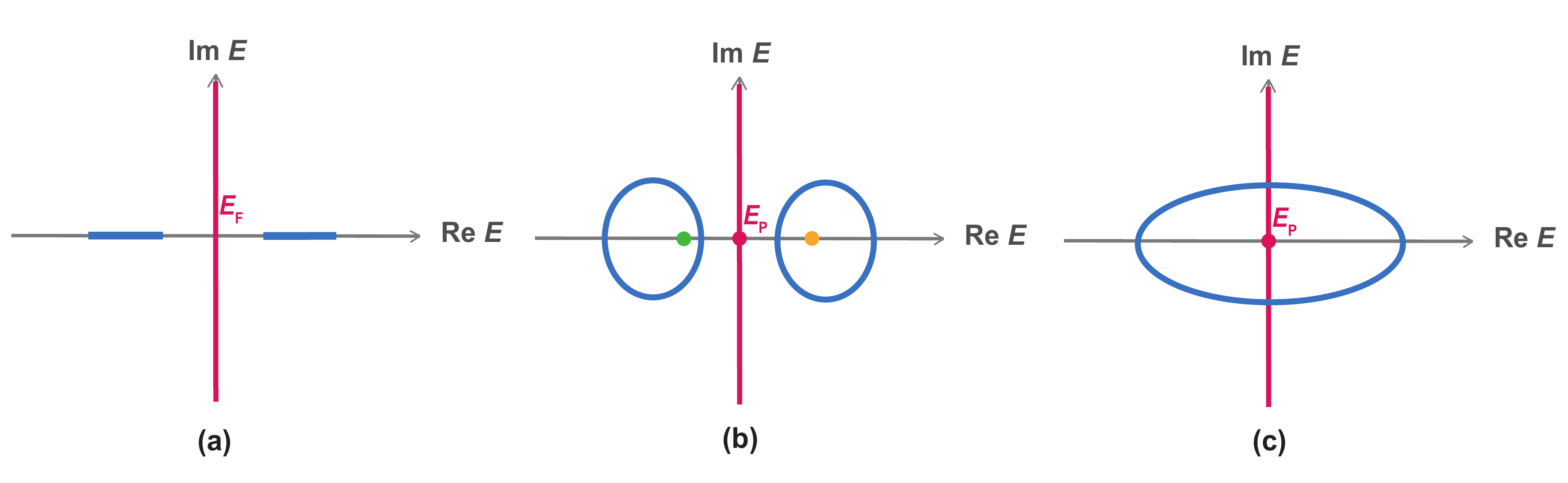}
    \caption{We show schematic illustrations of various types of spectral gaps in Hermitian and non-Hermitian quantum systems: (a) a gap (red line) of the real-valued spectrum (blue) of a Hermitian system, (b) the presence and (c) the absence of a line gap (red line) of the complex-valued spectrum (blue) of a non-Hermitian quantum system. The NHSE emerges due to the nonzero spectral winding and point-gap topology with respect to the reference point at $E_P=0$ (red dot) in (c) and at $E_P\neq 0$ (yellow and green dots) in (b). }
    \label{NHGap}
\end{figure*}

Non-Hermitian quantum systems are represented by non-Hermitian Hamiltonians $\hat{H}\neq \hat H^{\dagger}$, e.g., the well-known non-Hermitian Su-Schrieffer-Heegers (SSH) model: 
\begin{equation}
    \begin{aligned}
\hat{H}= & -\sum_{i}\left\{\left[1-(-1)^i \Delta J\right] c_{i}^{\dagger} c_{i+1}+H . c .\right\} \\
& +\delta \sum_{i}\left(c_{i}^{\dagger} c_{i+1}-c_{i+1}^{\dagger} c_{i}\right)-\mu\sum_i c^{\dagger}_ic_i,
\end{aligned}
\label{NHSSH}
\end{equation}
where $\mu$ is the Fermi energy, $\delta$ introduces non-reciprocal hopping thus non-Hermiticity, and $\Delta J$ describes a staggered hopping. For $\delta>0$ ($\delta<0$), the majority of the eigenstates become exponentially localized at the left (right) boundary under OBCs, a phenomenon known as the NHSE in sharp contrast to Hermitian counterparts where bulk eigenstates dominate. 

Given the complex spectrum, non-Hermitian quantum systems under PBCs may possess two types of energy gaps, protecting different topology: point gaps and line gaps. A point gap suggests that the spectrum does not cross a reference point $E_P$, i.e., $\operatorname{det}(\hat H-E_P) \neq 0$ \cite{origin2020}. Consequently, the spectral winding number around $E_P$:
\begin{equation}
    W(E_P)=\int_0^{2 \pi} \frac{d k}{2 \pi \mathrm{i}} \frac{d}{d k} \log \operatorname{det}\left[\hat{H}(k)-E_P \right],
    \label{bandwinding}
\end{equation}
remains topologically invariant under continuous variations as long as the point gap remains. When the winding number $W(E_P)\neq 0$ is nonzero around $E_P$, all line gaps across $E_P$ will close \cite{gong2018,kawabataprx,Chen2023real}; see Fig. \ref{NHGap}(c). In comparison, a line gap suggests that the complex spectrum never crosses a reference line [Fig. \ref{NHGap}(b)], whose physics and topological consequences are more analogous to gaps in Hermitian systems [Fig. \ref{NHGap}(a)]; here, the spectrum is fully separable into disjoint parts along such reference lines, and the winding number $W(E_P)$ will be zero. 

Importantly, the presence of spectral loops under PBC, i.e., a nontrivial spectral winding number $W(E_P)$ around arbitrary $E_P$, indicates the presence of the NHSE under OBC \cite{origin2020,zhang2020}. Therefore, according to the previous analysis on the point gap around $E_P$ and the line gap across $E_P$, the NHSE emerges in non-Hermitian systems under OBCs following the closure of line gaps under PBCs. On the other hand, an existing line gap does not guarantee the NHSE's absence, as there may exist nontrivial spectral winding with respect to other reference points $E_{P}$ [Fig. \ref{NHGap}(b)]. We may alter our reference point in the complex energy plane by varying the Fermi energy $\mu$ so that $E_P=\mu$ as we switch from an active perspective to a passive one. 

For example, the spectrum of the non-Hermitian SSH model retains a nontrivial spectral winding with respect to $E_P=0$ for $|\Delta J|<|\delta|$, forbidding a line gap through $E_P$ as illustrated in Fig. \ref{NHGap}(c); on the contrary, when $|\Delta J|>|\delta|$, a line gap develops in the spectrum of the non-Hermitian SSH model as shown in Fig. \ref{NHGap}(b), and the spectral winding with respect to $E_P=0$ vanishes. The transition happens at $|\Delta J|=|\delta|$ where the separate spectral loops meet and the line gap collapses. Also, for a variable $E_P$ at $|\Delta J|>|\delta|$, we may observe alternating point-gap topology and line gap, e.g., depending on the value of $E_P$ on the real axis [Fig. \ref{NHGap}(b)]. Also, similar to the insulator (gapped) and metal (gapless) phases of Hermitian systems, non-Hermitian systems may exhibit distinctive entanglement-entropy behaviors with the existence or absence of line gaps.

\subsection{SSE-QMC applicability towards non-Hermitian quantum systems}\label{sec:IIC}

For a non-Hermitian Hamiltonian $\hat{H}\neq \hat H^{\dagger}$, its right eigenstates $|\Psi_i^R\rangle$ and left eigenstates $|\Psi_i^L\rangle$ corresponding to eigenvalue $E_i$ \cite{Brody_2014}:
\begin{eqnarray}
\hat H|\Psi_i^R\rangle&=&E_i|\Psi_i^R\rangle, \nonumber\\
\hat H^{\dagger}|\Psi_i^L\rangle&=&E_i^{*}|\Psi_i^L\rangle,
\end{eqnarray}
are different in general, and obey the biorthogonal conditions $\langle\Psi_n^L|\Psi_m^R\rangle=\delta_{mn}$, $\hat{I}=\sum_n |\Psi_n^R\rangle\langle\Psi_n^L|$ instead. 

Given a non-Hermitian Hamiltonian in a biorthogonal form \cite{Brody_2014}:
\begin{equation}
\hat H=\sum_i E_i|\Psi_i^R\rangle\langle\Psi_i^L|,
\end{equation}
we note its partition function: 
\begin{eqnarray}
Z&=&\sum_n e^{-\beta E_n} =\sum_n e^{-\beta E_n}\langle\Psi_n^L|\sum_{\alpha}|\alpha\rangle\langle\alpha|\Psi_n^R\rangle \\\nonumber
&=&\sum_{\alpha} \langle\alpha|e^{-\beta \hat H}\sum_n |\Psi_n^R\rangle\langle\Psi_n^L|\alpha\rangle =\sum_{\alpha} \langle\alpha|e^{-\beta \hat H}|\alpha\rangle,
\end{eqnarray}
retains the definition under an orthogonal basis $\left\{|\alpha\rangle\right\}$ in Eq. \ref{HermitianPartition}. Therefore, the non-Hermiticity and biorthogonality of non-Hermitian quantum systems do not pose direct obstacles to the SSE-QMC method. 

Similar to the Hermitian cases, we require the matrix elements $\langle\alpha_p|\hat H_{a_p,b_p}|\alpha_{p-1}\rangle$ to be non-negative or the number of negative matrix elements to be even, so that the overall sampling probability remains positive-semidefinite (sign-problem-free) in SSE-QMC calculations. Such requirements mainly depend on the model parameters and operators rather than the Hermiticity. However, unlike the Hermitian cases, where the partition function is always real and positive, here, a positive-definite partition function is a requirement. For certain non-Hermitian systems \cite{Chen2023real}, e.g., $\hat H$ with $\mathcal{P} \mathcal{T}$ symmetry, the spectrum is either real or in complex-conjugate pairs, and the corresponding partition function is guaranteed to be real \cite{Meisinger_2013}: 
\begin{equation}
    \begin{aligned}
        Z&=\operatorname{Tr}\left[\sum_i e^{-\beta  E_i|\Psi_i^R\rangle\langle\Psi_i^L|}\right]\\
        &=\sum_{E_i \in \text{real}}e^{-\beta E_i}+\sum_{E_i \in \text{complex}}(e^{-\beta E_i}+e^{-\beta E_i^{*}}).
    \end{aligned}
\end{equation}
Therefore, although non-Hermitian quantum systems may possess potentially complex spectra, the SSE-QMC method is still viable as long as its matrix elements are sign-problem-free. Like in the Hermitian cases, the SSE-QMC method is an efficient and straightforward algorithm applicable to relatively large systems and even higher dimensions. 

We can also use the SSE-QMC method to study the ground-state properties of non-Hermitian quantum many-body systems. Here, we define the ground state as the eigenstate ($|\Psi_0^R\rangle$ and $|\Psi_0^L\rangle$) with the lowest real part of its eigenenergy. For a sufficiently low temperature (large $\beta$, e.g., $\beta=100$ in unit of common model parameters):
\begin{equation}
\langle\hat{A}\rangle_{LR} = \operatorname{Tr}\left[\hat{A}\sum_i e^{-\beta  E_i|\Psi_i^R\rangle\langle\Psi_i^L|}\right]/Z \approx
\langle\Psi_0^L|\hat{A}|\Psi_0^R\rangle. 
\end{equation}

\subsection{Example: non-Hermitian quantum spin chains}\label{sec:IID}

Without loss of generality, let us consider the following non-Hermitian quantum spin chain of length $N$:
\begin{eqnarray}
    \hat H &=& \sum_b J_z S^z_b S^z_{b+1} + [1- (-1)^b \Delta J] (S^x_b S^x_{b+1}+S^y_b S^y_{b+1}) \nonumber \\
    & & + i\delta (S^x_b S^y_{b+1}-S^y_b S^x_{b+1}) \nonumber\\
    &=&\sum_b J_z S^z_b S^z_{b+1}+\frac{1}{2}[1-(-1)^{b}\Delta J-\delta] S_{b}^{+} S_{b+1}^{-} \nonumber \\
    & & + \frac{1}{2}[1-(-1)^{b}\Delta J+\delta] S_{b}^{-} S_{b+1}^{+}, 
    \label{eq:spinchain}
\end{eqnarray}
where $J_z, \Delta J, \delta \in \mathbb{R}$ are model parameters: $J_z$ is an Ising-type interaction, $\Delta J$ is a staggered $XY$ interaction, and $\delta$ is responsible for the overall non-Hermiticity of the model. For OBC, the summation of $b$ runs between 1 and $N-1$, while we sum over $b\in [1, N]$ and identify $b=1, N+1$ for PBC. The model is $\mathcal{P} \mathcal{T}$ symmetric, so we can feel free to use SSE-QMC here.

To apply the SSE-QMC method, we decompose the Hamiltonian as:
\begin{equation}
\begin{aligned}
    \hat H&=-\sum_b \hat H_{1,b}- \hat H_{2, b}- \hat H_{3, b},\\
    \hat H_{1, b}&=C-J_z S_{b}^z S_{b+1}^z,\\
    \hat H_{2, b}&=\frac{1}{2}[1-\Delta J(-1)^{b}-\delta] S_{b}^{+} S_{b+1}^{-},\\
    \hat H_{3, b}&=\frac{1}{2}[1-\Delta J(-1)^{b}+\delta] S_{b}^{-} S_{b+1}^{+},\\
\end{aligned}
\end{equation}
where $C=\epsilon+J_z/4$ is a constant that alters some matrix elements while keeping the model physics invariant. We also regard $\hat H_{2,b}$ and $\hat H_{3,b}$ as two separate off-diagonal operators. Their coefficients differ when $\delta \neq 0$ and allow $\hat{H}$ to be non-Hermitian. Correspondingly, the partition function takes the following form:
\begin{equation}
    Z=\sum_{\alpha,S_n}\frac{\beta^n}{n!}(-1)^{n_2+n_3} \langle\alpha|\prod_{p=1}^n \hat H_{a_p,b_p}|\alpha\rangle,
\end{equation}
where $n_2$ and $n_3$ are the number of $\hat{H}_{2,b}$ and $\hat{H}_{3,b}$ operators in the operator sequence $\{[a_p, b_p]\}$, respectively. For a quantum spin chain with an even number $N$ of sites, the total number of off-diagonal operators that shift a spin up by one lattice spacing, $n_2+n_3$, is always even irrespective of the configurations. Thus, we can safely drop the $(-1)^{n_2+n_3}$ factor. The nonzero matrix elements of the nontrivial operators are:
\begin{equation}
    \begin{aligned}
        W_{11}&=\langle\uparrow\uparrow|\hat H_{1,b}|\uparrow\uparrow\rangle=\epsilon,\\        W_{12}&=\langle\downarrow\downarrow|\hat H_{1,b}|\downarrow\downarrow\rangle=\epsilon,\\        W_{13}&=\langle\uparrow\downarrow|\hat H_{1,b}|\uparrow\downarrow\rangle=\epsilon+J_z/2,\\       W_{14}&=\langle\downarrow\uparrow|\hat H_{1,b}|\downarrow\uparrow\rangle=\epsilon+J_z/2,\\        W_2&=\langle\uparrow\downarrow|\hat H_{2,b}|\downarrow\uparrow\rangle=\frac{1}{2}[1-\Delta J(-1)^{b}-\delta],\\        W_3&=\langle\downarrow\uparrow|\hat H_{3,b}|\uparrow\downarrow\rangle=\frac{1}{2}[1-\Delta J(-1)^{b}+\delta],\\ 
    \end{aligned} \label{eq:Ws}
\end{equation}
whose vertices are illustrated in Fig. \ref{vertexandConfig}(a). To meet the positive-semidefinite requirement on such vertices, we should make the model parameters satisfy $1-|\Delta J|-|\delta|\geq0$ and $\epsilon\geq\operatorname{max}(0,-J_z/2)$. The resulting model is sign-problem-free for the SSE-QMC method. 

\begin{figure}
    \centering
    \includegraphics[width = 1.1\linewidth]{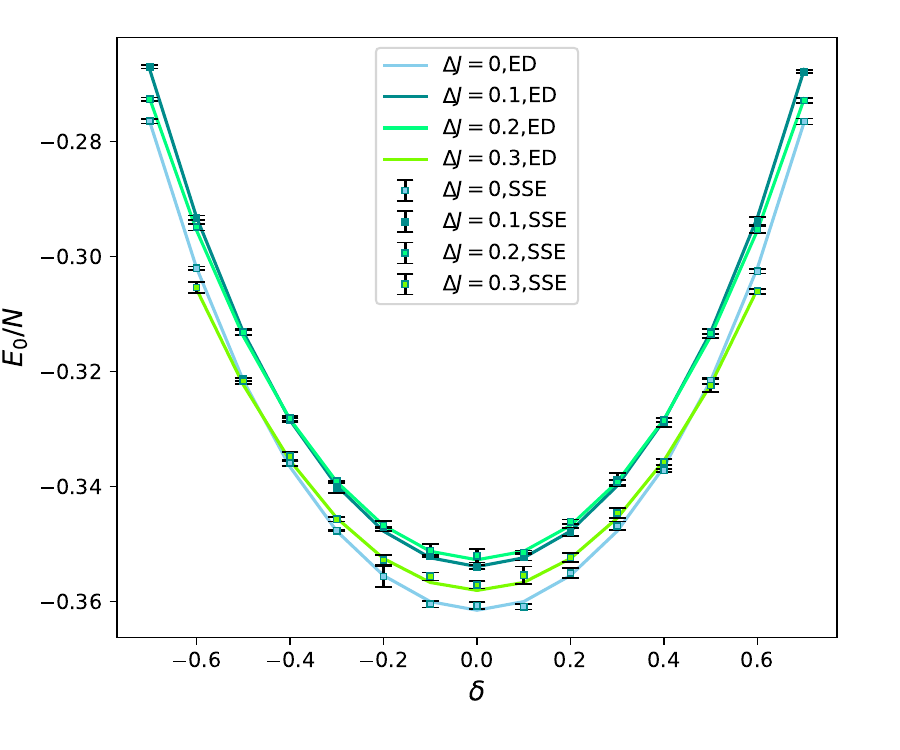}
    \caption{The energies of the non-Hermitian quantum spin chain in Eq. \ref{eq:spinchain} for various $\delta$ and $\Delta J$ under OBCs compare consistently between ED ground states and SSE-QMC calculations at low temperature $\beta=100$. $J_z=0.5$ and $N=12$.}
    \label{OBCE0_delta}
\end{figure}

For benchmark, we first calculate the energies of non-Hermitian quantum spin chains under OBCs at a low temperature $\beta=100$ with the SSE-QMC method and compare with the ground-state energy via exact diagonalization (ED) for relatively small systems $N=12$. The ED results have also confirmed that the models host real spectra, which pose no problem for the SSE-QMC method. We summarize the results for various $\delta$ and $\Delta J$ with a finite $J_z=0.5$ in Fig. \ref{OBCE0_delta}, showing satisfactory consistency and that SSE-QMC works well on non-Hermitian systems with interactions. 

\begin{figure}
    \centering
    \includegraphics[width = 1.1\linewidth]{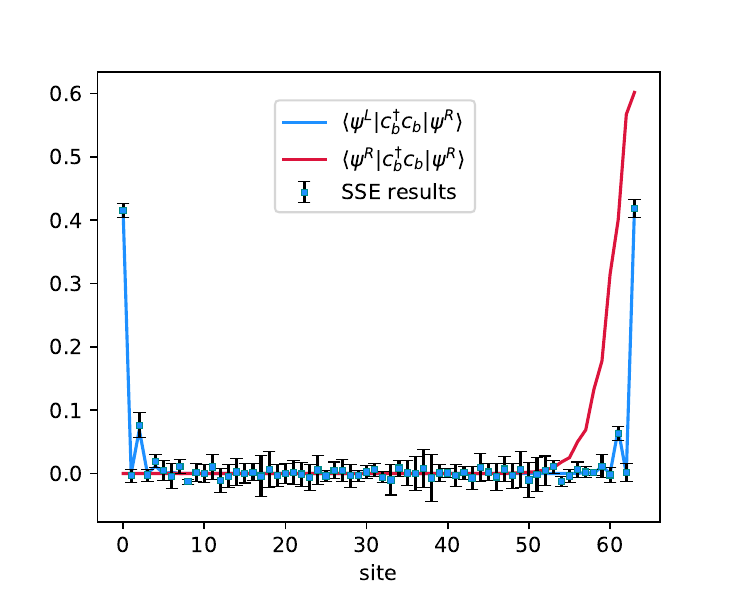}
    \caption{We can map a $J_z=0$ quantum spin chain to a free-fermion model through the Jordan-Wigner transformation. We evaluate the density distribution of a target single-particle state via (1) the modulus square of the right eigenstate $|\psi^R\rangle$ of the free-fermion model, (2) the biorthogonal expectation value $\langle \psi^ L|c^\dagger_b c_b |\psi^R\rangle$ of the free-fermion model, and (3) the difference between densities of the $S_z^{tot}=0, -1$ sectors in the SSE-QMC calculations at low-temperature $\beta=100$ - these two sectors differ by the participation or the absence of the single-particle state $|\psi^R\rangle$ (and $\langle \psi^ L|$). The NHSE is only present in the first case \cite{PhysRevB.101.121109}, and the latter two cases show consistent expectation values under the biorthogonal basis and present edge effects rather than the NHSE. We set $\delta=0.5$ and $\Delta J=0.3$ to trigger the NHSE for $N=64$ under OBCs. }
    \label{OBC_dis_delta_Large}
\end{figure}

Interestingly, we can map the non-Hermitian quantum spin chain in Eq. \ref{eq:spinchain} to a non-Hermitian interacting fermion chain through the Jordan-Wigner transformation \cite{Nagaosa1999QuantumFT}:
\begin{eqnarray}
    S^z_i&=&f_i^{\dagger}f_i-\frac{1}{2}, \nonumber\\
    S_i^{+}S_{i+1}^{-}&=&f_i^{\dagger}f_{i+1}, \nonumber\\
    S_i^{-}S_{i+1}^{+}&=&f_{i+1}^{\dagger}f_{i}, \label{eq:jordan}
\end{eqnarray}
where a spin-up (spin-down) site in the spin model corresponds to an occupied (empty) site in the fermion model. Likewise, the worldlines trace the fermions and form closed loops in the fermion model. Therefore, the SSE-QMC method also generalizes straightforwardly to non-Hermitian interacting fermion systems. 

In particular, the corresponding fermion chain is non-interacting when $J_z=0$. We note that the single-particle right eigenstates of non-Hermitian free-fermion chains may exhibit the NHSE, as shown in Fig. \ref{OBC_dis_delta_Large}. However, the NHSE is absent from the quantum many-body perspective, as forbidden by the Pauli exclusion principle \cite{PhysRevB.101.121109} and under the biorthogonal basis. Indeed, we evaluate the density distribution of a single-particle state by taking the difference between two many-body densities with $n_f=N/2, N/2-1$ fermions ($S_z^{tot}=0, -1$ under the quantum spin representation) with or without the target single-particle state, respectively. The results display no NHSE and are consistent with the density expectation values under the biorthogonal basis; see Fig. \ref{OBC_dis_delta_Large}. Such consistency also indicates that our SSE-QMC calculations are readily applicable to relatively large systems and low temperatures.

\section{Nontrivial worldline winding in non-Hermitian quantum systems}\label{sec:III}

\subsection{QMC difficulty for non-Hermitian systems under PBC}\label{sec:IIIA}

\begin{figure}
    \centering
    \includegraphics[width = 1\linewidth]{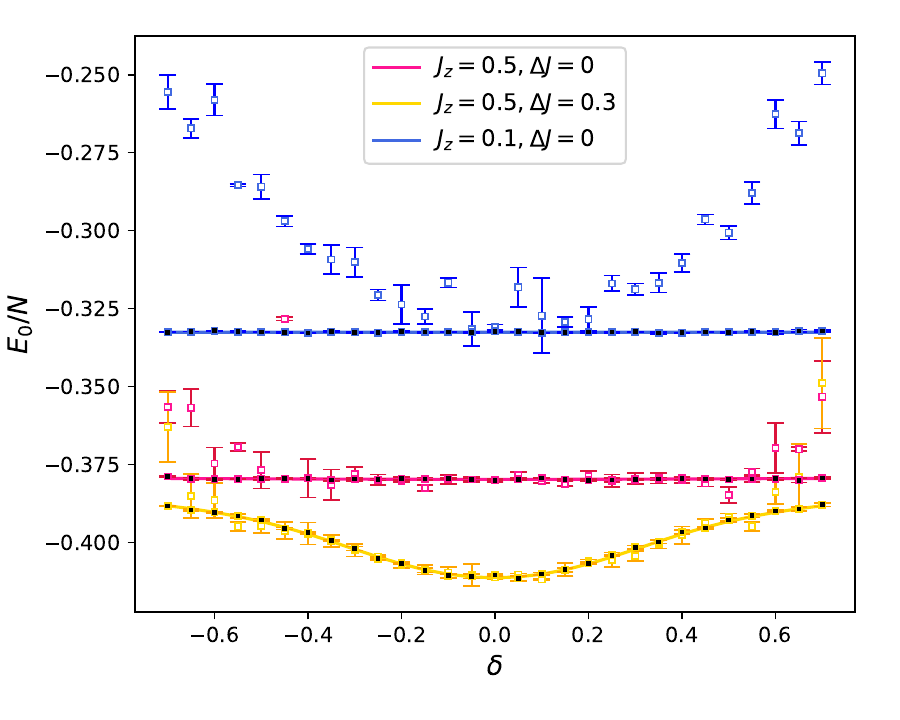}
    \caption{While the SSE-QMC results with $\epsilon=0$ (hollow points) deviate from the ED benchmark (solid curves) beyond uncertainties, especially for larger non-Hermitian parameter $\delta$, they fit well with $\epsilon=0.5$ (black points). We consider non-Hermitian quantum spin chains under PBCs with various $\delta$, $\Delta J$, and $J_z$. $\beta=100$ and $N=12$. }
    \label{PBCE0_delta}
\end{figure}

Unlike the OBC cases, however, the SSE-QMC calculations for non-Hermitian models under PBC sometimes strike obstacles and fail to converge to the benchmark values. For example, we evaluate the ground-state energies of various non-Hermitian quantum spin chains under PBCs, and the divergences between SSE-QMC results and ED benchmarks are clearly beyond an uncertainty explanation; see Fig. \ref{PBCE0_delta}. Such deviation generally increases with the non-Hermitian parameter $\delta$ and decreases with $\Delta J$ and $J_z$. We will first give a prompt answer on the origin of such difficulty; in later subsections, we will give more quantitative studies and discuss its possible resolution and physical consequences. 

\begin{table}
    \centering
    \begin{tabular}{c|c|c|c|c|c|c|}
         $\epsilon=0$ & $w=0$ & $w=1$ & $w=2$ & $w=3$ & $w=4$ & $w=5$  \\
         \hline
         $N_{w}$ & 578 & 903 & 871 & 64444 & 285491 & 571172 \\
         \hline
         $N_{\Delta w<0}$ & 0 & 0 & 0 & 0 & 0 & 0 \\
         \hline
         $N_{\Delta w>0}$ & 1 & 1 & 1 & 1 & 1 & 2 \\
         \hline
         $\epsilon=0$ & $w=6$ & $w=7$ & $w=8$ & $w=9$ & $w=10$ & $w=11$\\
         \hline
         $N_{w}$ & 553020 & 496057 & 1710886 & 313890 & 2732 & 0\\
         \hline
         $N_{\Delta w<0}$ & 1 & 1 & 6 & 2 & 1 & 0\\
         \hline
         $N_{\Delta w>0}$ & 2 & 6 & 2 & 1 & 0 & 0\\
         \hline
         $\epsilon=0.5$ & $w=3$ & $w=4$ & $w=5$ & $w=6$ & $w=7$ & $w=8$  \\
         \hline
         $N_{w}$ & 767 & 5663 & 28225 & 107078 & 287879 & 572401 \\
         \hline
         $N_{\Delta w<0}$ & 8 & 96 & 630 & 2899 & 9516 & 22820 \\
         \hline
         $N_{\Delta w>0}$ & 114 & 744 & 3280 & 10601 & 24412 & 40677 \\
         \hline
         $\epsilon=0.5$ & $w=9$ & $w=10$ & $w=11$ & $w=12$ & $w=13$ & $w=14$\\
         \hline
         $N_{w}$ & 825230 & 879035 & 679295 & 385448 & 163468 & 51300\\
         \hline
         $N_{\Delta w<0}$ & 39274 & 49521 & 44386 & 28973 & 13948 & 4835\\
         \hline
         $N_{\Delta w>0}$ & 49566 & 43286 & 27412 & 12809 & 4287 & 1045\\
         \hline
    \end{tabular}
    \caption{The number of samples $N_{w}$ with worldline winding number $w$ and the number of updates $N_{\Delta w>0}$ ($N_{\Delta w<0}$) with increasing (decreasing) $w$ in a typical SSE-QMC trial show the limited (enhanced) ergodicity for $\epsilon=0$ ($\epsilon=0.5$). We consider a non-Hermitian quantum spin chain with $J_z=0.1$, $\Delta J=0$, $\delta=0.3$, $\beta=100$ and $N=10$ under PBCs. For $\epsilon=0$, $N_{\Delta w>0}$ and $N_{\Delta w<0}$ are extremely small, limiting the sample distribution $N_{w}$ from reaching ergodicity. In contrast, the transition between $w$ is much more fluent for $\epsilon=0.5$, leading to a sample distribution $N_{w}$ concentrated around the dominant winding number $w_{opt}\neq 0$.}
    \label{tab:ergodicity}
\end{table}

In Sec. \ref{sec:IIA}, we discussed the concept of worldlines in $(D+1)$-dimensional space-time and their corresponding winding number $w$. Obviously, we have $w=0$ in the cases of OBCs; under PBCs, however, worldlines may possess nontrivial winding numbers $w\neq 0$, i.e., wrap around the system along a periodic spatial direction for a finite number of net times before returning to the initial spot as it evolves under imaginary time. Indeed, the problem in SSE-QMC calculations for non-Hermitian quantum systems under PBCs is associated with such global loops and winding numbers: (1) the dominant worldline sector in the partition function, thus in the SSE-QMC sampling, may shift to $w_{opt}\neq 0$, and (2) the transitions between different winding-number sectors are limited, breaking the ergodicity essential for convergence; see Table \ref{tab:ergodicity} for example.

\subsection{Nontrivial worldline winding from a non-Hermitian toy-model perspective}\label{sec:IIIB}

To illustrate such a nontrivial distribution of worldline winding numbers in non-Hermitian quantum systems, we consider the following non-Hermitian toy model on a 1D periodic system ($\alpha \in \mathbb{R}$):
\begin{equation}
    \hat{H}=-\frac{\partial^2}{\partial{\theta}^2}+\alpha\frac{\partial}{\partial{\theta}},
    \label{NH_on_a_Ring}
\end{equation}
whose eigenstates \footnote{The left and right eigenstates are identical in this case.} and eigenenergies are:
\begin{equation}
    \psi_m(\theta) =\exp (i m \theta),\qquad E_m = m^2+i\alpha m, 
    \label{NH_on_a_Ring_eigen}
\end{equation}
where $m\in \mathbb{Z}$ is the angular momentum. 

Following the imaginary-time path-integral formalism, we can derive the partition function as:
\begin{eqnarray}
Z&=&\int D \theta \prod_{j=1}^N\left\langle\theta_{j+1}|\exp (-\Delta \tau H)| \theta_j\right\rangle \label{PartitionFunc} \\\nonumber
&=&\int D \theta \prod_{j=1}^N \Big\{\sum_{m_l} \exp \left[i m_l\left(\theta_{j+1}-\theta_j\right) -\Delta \tau\left(m_l^2+i \alpha m_l\right)\right]\Big\} \\\nonumber
&=&\int D \theta \prod_{j=1}^N\left\{\sum_{n_l} \exp \left[-\frac{\left(\theta_{j+1}-\theta_j+2 \pi n_l-\alpha\,\Delta\tau\right)^2}{4 \Delta \tau}\right]\right\},
\end{eqnarray}
where $\Delta \tau=\beta / N$ is a small discrete step in the imaginary-time direction, labeled by $j$, with $\theta_{N+1}=\theta_1$. We have employed Poisson’s summation formula in the last line. 

To tackle such a functional integral, we start from a typical path:
\begin{equation}
    \theta_j=\bmod \left[\theta_1+\frac{2 \pi w}{N}(j-1)+\delta \theta_j, 2 \pi\right],
\end{equation}
where $w$ is $\theta$'s winding number and $\delta \theta_j$ are local fluctuations that are essentially independent of $w$: 
\begin{equation}
    \theta_{j+1}-\theta_j=\left\{
\begin{aligned}
\frac{2\pi w}{N} & , & \theta_{j}+\frac{2\pi w}{N}<2\pi, \\
\frac{2\pi w}{N}-2\pi & , & \theta_{j}+\frac{2\pi w}{N}>2\pi,
\end{aligned}
\right.
\end{equation}
where $\theta$ goes across $2\pi$ from $j$ to $j+1$ for the second line. Consequently, for $\Delta\tau\rightarrow 0$, i.e., $N\rightarrow\infty$, the summation over $n$ in Eq. \ref{PartitionFunc} is dominated by $n=0$ so that $\left(\theta_{j+1}-\theta_j+2 \pi n-\alpha\,\Delta\tau\right)^2\approx0$, unless $\theta$ goes across $2\pi$ from $j$ to $j+1$, where $n=1$ dominates. As a result, after keeping only the contributing terms, we obtain:
\begin{equation}
    \begin{aligned}
        Z&=f(\beta)\sum_{w=-\infty}^{+\infty}\prod_{j=1}^N\mathrm{exp}\left\{-\frac{(2\pi w-\alpha\beta)^2}{4N\beta}\right\}\\
        &=f(\beta)\sum_{w=-\infty}^{+\infty}\mathrm{exp}\left[-\frac{(2\pi w-\alpha\beta)^2}{4\beta}\right],\\
    \end{aligned}
    \label{Partition_w}
\end{equation}
where $f(\beta)$ is a function on the effects of $\delta\theta_j$ fluctuations independent of $w$. 

The partition function in Eq. \ref{Partition_w} characterizes the weights and importance of different winding-number sectors, which contain imaginary-time path-integral worldlines that wrap around the $[0,2\pi]$ interval a net $w$ number of times. In an ideal QMC sampling process, the larger the weight of a particular winding number $w$, the more frequently we should sample the corresponding sector's configurations. For the Hermitian case with $\alpha=0$, the partition function is dominated by the $w=0$ sector \cite{PhysRevB.57.13382}. Especially, the weights for different winding numbers converge at low temperatures (large $\beta$); thus, calculations in a specific sector, e.g., $w=0$ for typical initializations, are as good as calculations that run through all sectors \cite{PhysRevB.57.13382}. However, for the non-Hermitian cases $\alpha\neq0$, the worldline configurations with nontrivial winding $w_{opt}=\alpha\beta/2\pi$ have the largest weight. Moreover, the location of the most probable sector moves farther away from $w=0$ as $\beta$ increases. As a result, we need to ensure that all sectors, if not the sectors around $w_{opt}=\alpha\beta/2\pi$ in particular, are appropriately represented in the sampling and calculations. 

However, such worldline winding numbers are essentially topological quantities, and updates that alter winding are scarce and rarely accepted in SSE-QMC calculations. Consequently, we may encounter a problem with ergodicity: the configurations are stuck near the initial $w$ far away from $w_{opt}$, leading to incomprehensive sampling and, therefore, inaccurate evaluations, as we demonstrated in Sec. \ref{sec:IIIA} and Table \ref{tab:ergodicity}. For more ergodic SSE-QMC calculations, we may introduce a remedy by enhancing the transition rates between different worldline winding-number sectors, which we discuss next.

\subsection{Enhanced ergodicity between winding-number sectors}\label{sec:IIIC}

To enhance the ergodicity between different winding-number sectors, we dig into the proposed updates in the directed loop update algorithm. The vertices are at the center of the proposed updates to the worldlines. There are four possible legs for the exit given an entrance leg into a vertex; if the exit and entrance legs are identical, the proposed loop experiences a bounce process \cite{PhysRevE.66.046701}. Intuitively, we wish to minimize or at least reduce the bounce probability to allow the loop to propagate and proliferate and end up with more global loops so that they may alter the winding number more efficiently. However, we do not have many  degrees of freedom for maneuvering: parameters like $N$, $\beta$, $J_z$, $\delta$, and $\Delta J$ are all physically relevant. Fortunately, there are model-independent parameters, such as $\epsilon$, which we can tune to adjust the bounce probability and enhance ergodicity without causing changes in physics. 

Without loss of generality, we consider vertex $W_3$ with the entrance leg in the lower left as an example, whose probability of updated vertex with corresponding exit leg is: 
\begin{equation}
    P(W_3 \rightarrow W_j)=\frac{W_j}{W_{11}+W_{13}+W_3}, \label{vertexprob_W4}
\end{equation}
where $W_{11}$, $W_{13}$, and $W_3$ are the (weights of) vertices associated with the exit legs in the upper right, the upper left, and the lower left, respectively (the exit leg in the lower right has no corresponding vertex and thus zero matrix element); see Fig. \ref{vertexandConfig}a and Eq. \ref{eq:Ws}. In particular, the probability of the bounce process, where the exit and entrance legs are identical, and the vertex remains unchanged, is: 
\begin{equation}
P_{\mathrm{bounce}}=P(W_3 \rightarrow W_3)=\frac{1-(-1)^b \Delta J+\delta} {\left[1-(-1)^b \Delta J+\delta \right]+4\epsilon+J_z}. 
\end{equation}
Therefore, we can reduce the bounce probability by increasing $\epsilon$. 

\begin{figure}
    \centering
    \includegraphics[width = 1\linewidth]{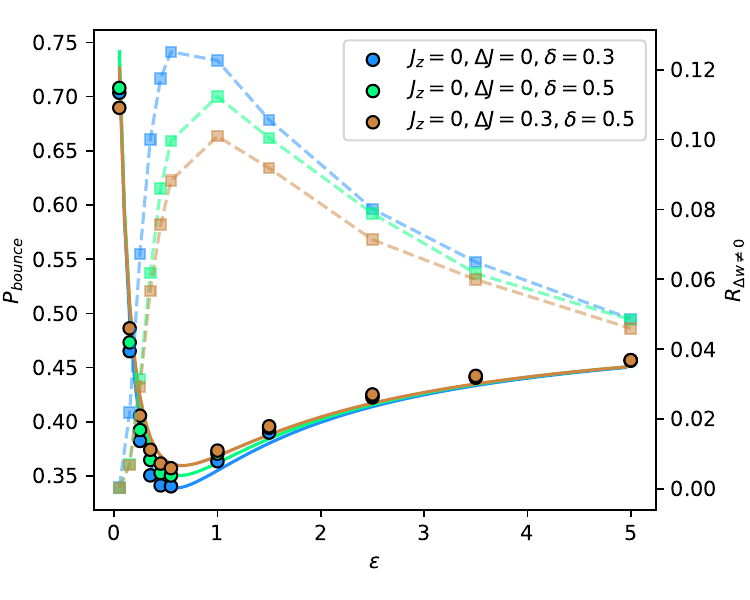}
    \caption{The average bounce probabilities $P_{bounce}$ in the SSE-QMC loop updates (solid circles) and the semi-quantitative analysis (solid curves) consistently indicate the effect of a finite $\epsilon$ in lowering $P_{bounce}$ and in turn, elevating the ratio $R_{\Delta w\neq 0}$ of winding-number-altering updates (dashed lines) thus the overall ergodicity in the SSE-QMC processes. Optimal $R_{\Delta w\neq 0}$ reaches $>8\%$ for $P_{bounce}$ around $\epsilon \in [0.5, 1]$. We consider non-Hermitian models for various values of $J_z$, $\Delta J$, and $\delta$ with $\beta=100$ and $N=12$ under PBCs. }
    \label{avgbounce}
\end{figure}

More comprehensively, we may estimate the average bounce probability semi-quantitatively as follows. As we discussed in Sec. \ref{sec:IIA}, we can relate the operator expectation values $\langle \hat H_{a,b}\rangle=\langle n_{a,b}\rangle/\beta$ with their (average) instances $\langle n_{a,b}\rangle$ appearing in the operator sequence $S_M$. Therefore, we have:
\begin{equation}
    \begin{aligned}
       \langle n_{1,b}\rangle/\beta= \langle \hat H_{1,b}\rangle&=\epsilon+J_z/4-J_z\langle S_{b}^zS_{b+1}^z\rangle,\\
       \langle n_{2,b}\rangle/\beta= \langle \hat H_{2,b}\rangle&=\frac{1}{2}(1-\delta-\Delta J(-1)^b)\langle S_{b}^+S_{b+1}^-\rangle,\\
       \langle n_{3,b}\rangle/\beta= \langle \hat H_{3,b}\rangle&=\frac{1}{2}(1+\delta-\Delta J(-1)^b)\langle S_{b}^-S_{b+1}^+\rangle.\\
    \end{aligned} \label{eq:n123}
\end{equation}
Further, we can divide $\langle n_{1,b}\rangle$ of the diagonal operator $\hat H_{1,b}$ into that of its four vertices: $\langle n_{11,b}\rangle / \epsilon = \langle n_{12,b}\rangle /\epsilon = \langle n_{13,b}\rangle / (\epsilon+J_z/2)= \langle n_{14,b}\rangle / (\epsilon + J_z/2)$ following Eq. \ref{eq:Ws}. As a result, we can roughly establish the ratio of each type of vertices in SSE-QMC samples from the correlation functions:
\begin{equation}
    \begin{aligned}
         \frac{\langle n_{11,b}\rangle}{\beta} = \frac{\langle n_{12,b}\rangle}{\beta}&=\frac{ 2\epsilon \left(\epsilon+J_z/4-J_z\langle S_{b}^zS_{b+1}^z\rangle \right)}{4\epsilon+J_z},\\
        \frac{\langle n_{13,b}\rangle}{\beta} = \frac{\langle n_{14,b}\rangle}{\beta}&=\frac{ \left(2\epsilon+J_z \right) \left(\epsilon+J_z/4-J_z\langle S_{b}^zS_{b+1}^z\rangle \right)}{4\epsilon+J_z}.\\
    \end{aligned} \label{eq:n11234}
\end{equation}
Then, we can estimate the bounce probability: 
\begin{equation}
    \Bar{P}_{\mathrm{bounce}}(i)=\sum_{j} \frac{ \langle n_{j,b}\rangle} { \sum_{j'} \langle n_{j',b}\rangle} P(W_j \rightarrow W_j), \label{eq:est_pb}
\end{equation}
by averaging over the vertices with respect to their weights in Eqs. \ref{eq:n123} and \ref{eq:n11234}. 

We summarize the bounce probability and the ratio $R_{\Delta w\neq 0}$ of worldline-winding-altering loops in directed loop updates among the SSE-QMC calculations for varying $\epsilon$ in Fig. \ref{avgbounce}. The semi-quantitative bounce probability in Eq. \ref{eq:est_pb} also presents a reasonable estimation. For $\epsilon=0$, the bounce probability is nearly 0.9, and $R_{\Delta w\neq 0}$ is nearly zero, hampering effective transitions between different worldline winding-number sectors; in comparison, the bounce probability drops below 0.4 for $\epsilon \in [0.5, 1.0]$, and subsequently, $R_{\Delta w\neq 0}$ approaches nearly $10\%$, providing enhanced ergodicity in SSE-QMC sampling. 

\begin{figure}
    \centering
    \includegraphics[width=1.05\linewidth]{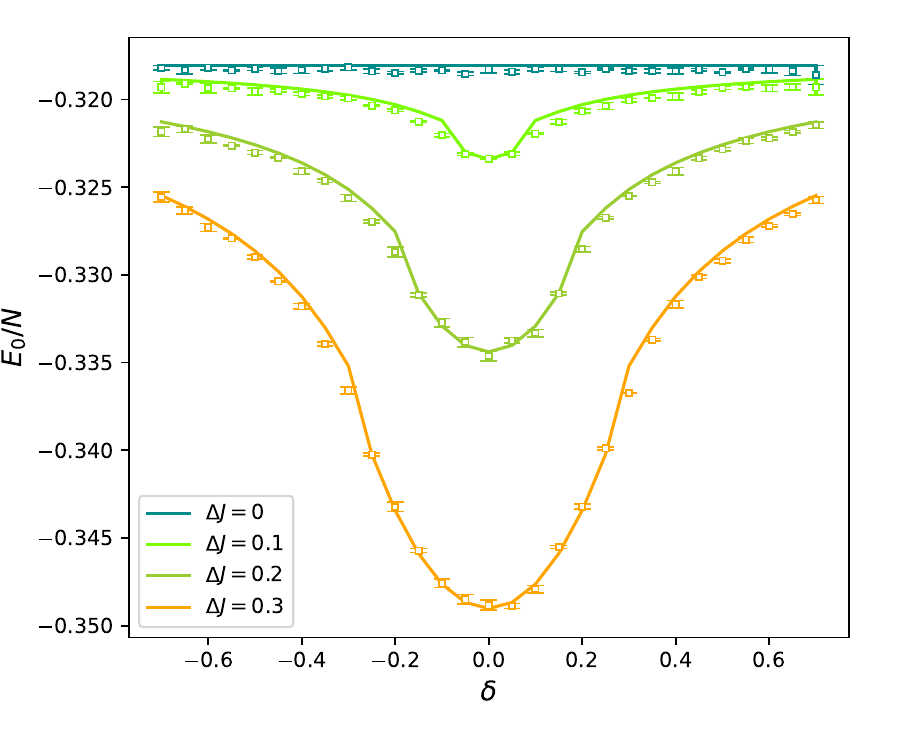}
    \caption{With finite $\epsilon=0.5$, the energies of the non-Hermitian quantum spin chains ($J_z=0$) from the SSE-QMC calculations at low temperature $\beta=100$ (square points) compare very well with theoretical ground-state values (solid curves) obtained through the Jordan-Wigner transformation for various $\delta$, $\Delta J$, and relatively large systems $N=64$ under PBCs. }
    \label{PBC_Energy_withbar}
\end{figure}

Indeed, introducing a finite $\epsilon=0.5$ enhances ergodicity under PBCs and yields consistent results in the SSE-QMC calculations. As summarized in Fig. \ref{PBCE0_delta}, the SSE-QMC results on non-Hermitian quantum spin chains witness satisfactory consistency with the ED benchmarks upon setting $\epsilon=0.5$, with remarkable improvements over and contrast with the $\epsilon=0$ results plagued by nontrivial worldline winding. Such characteristic disparities in efficiency on changing winding numbers are also apparent in Table \ref{tab:ergodicity}. The remedy also works on relatively large non-Hermitian quantum systems, where, with global and topological distinctions, the barrier between different worldline winding sectors and the ergodicity issue is intuitively more severe. For instance, we compare the SSE-QMC results for $J_z=0$ under PBCs with the non-Hermitian free-fermion models upon the Jordan Wigner transformation and obtain consistent results on relatively large systems ($N=64$), see Fig. \ref{PBC_Energy_withbar}, suggesting the nontrivial worldline winding no longer poses an apparent obstacle. We note that such a remedy is not unique or exclusive, as there exist other ways to enhance ergodicity between different winding-number sectors, such as periodically proposing updates that insert specific vertices leading to a new worldline with $\pm 1$ winding number. 


\subsection{Investigation on model conditions for nontrivial worldline winding}\label{sec:IIID}

\begin{figure}
    \centering
    \includegraphics[width = 1\linewidth]{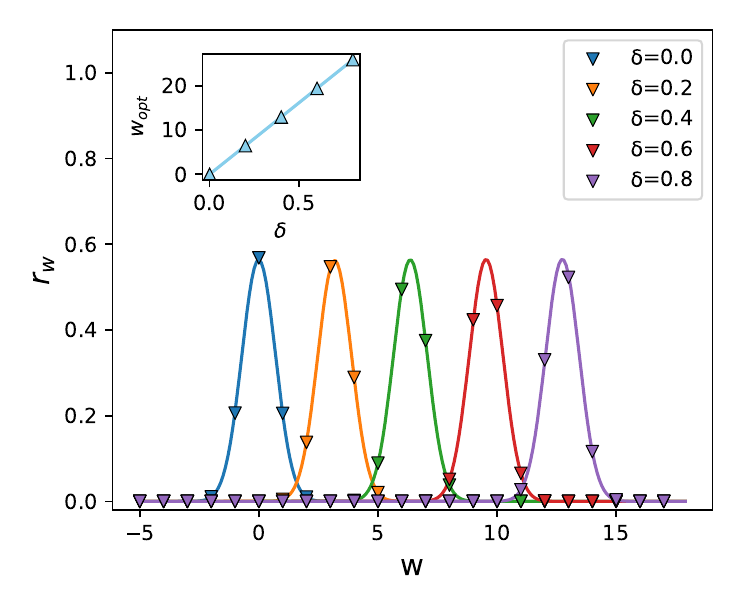}
    \caption{The (normalized) distribution of worldline winding numbers generally shifts from 0 towards larger values as the non-Hermitian parameter $\delta$ increases from 0. Inset: The Gaussian-fit peak positions $w_{opt}$ of the winding-number distributions follow a linear relation to $\delta$. Here, we set $J_z=0$ for free-fermion chains, $\Delta J=0$ for a closed line gap, $\epsilon=0.5$ for enhanced ergodicity, and $\beta=100$, $N=64$ under PBCs. }
    \label{vardelta}
\end{figure}

Previously in Sec. \ref{sec:IIIB}, we have shown in the toy model that, unlike Hermitian models, the most dominant worldline winding is no longer necessarily the $w=0$ sector in non-Hermitian quantum systems. Such nontrivial winding numbers may cause difficulty in ergodicity and deviations in expectation values (Sec. \ref{sec:IIIA}). Here, through numerical studies of various non-Hermitian quantum spin chains with PBCs and enhanced ergodicity (Sec. \ref{sec:IIIC}), we keep track of the worldline winding numbers $w$ during our SSE-QMC calculations and analyze the systematic conditions of such nontrivial worldline winding. Importantly, the conditions of nontrivial worldline winding coincide with nontrivial point-gap topology for the reference point $E_P=0$. 

We summarize the results for varying $\delta$ and fixed $J_z=\Delta J=0$ and $\beta=100$ in Fig. \ref{vardelta}. The resulting models are equivalent to free-fermion models with vanishing line gaps following the Jordan-Wigner transformation. Like the gapless non-Hermitian toy model in Eq.  \ref{NH_on_a_Ring}, the winding number distributions, normalized as a ratio $r_w=N_w/\sum_{w}N_w$, display a Gaussian-shaped pattern; notably, the fitted peak of the distribution sits at $w_{opt}=0$ for $\delta=0$ and gradually shifts to the right $w_{opt}>0$ as the non-Hermitian parameter $\delta$ - the amplitude difference between the right hopping $\hat H_{2,b}$ and the left hopping $\hat H_{3,b}$ - increases. Such linear relation is comparable to Eq. \ref{Partition_w} of the gapless toy model.

\begin{figure}
    \centering
    \includegraphics[width = 1.05\linewidth]{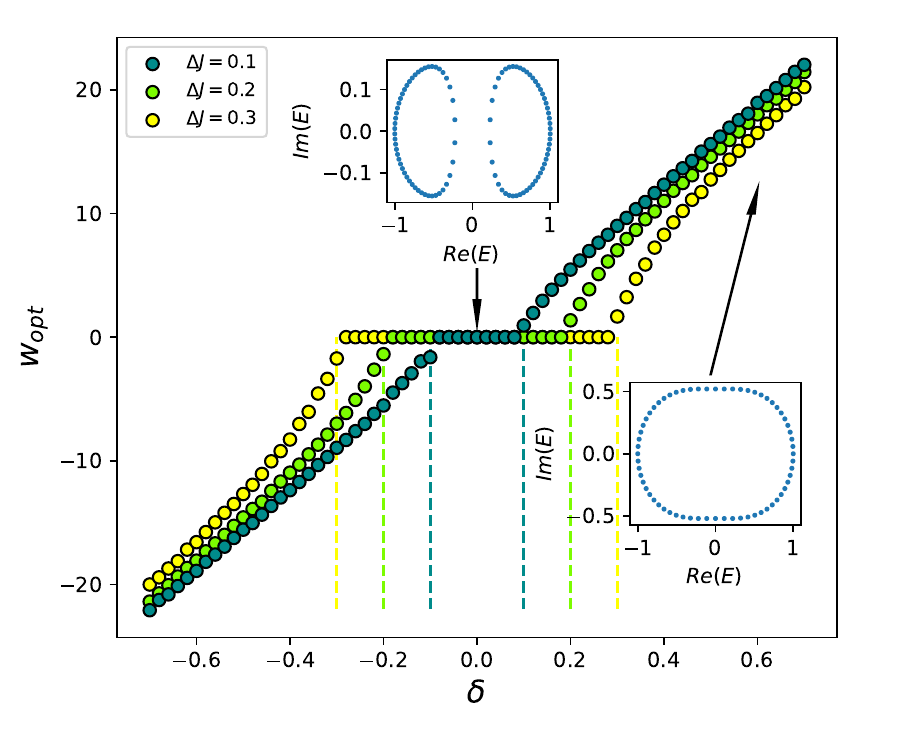}
    \caption{The dominant worldline winding number $w_{opt}$ deviates from zero as the non-Hermitian parameter $\delta$ surpasses the line-gap $\Delta J$ (dashed lines) and evolves monotonically afterward. $J_z=0$, $\beta=100$, $\epsilon=0.5$, and $N=64$ under PBCs. The parameter space for nonzero (zero) $w_{opt}$ is consistent with the presence (absence) of non-Hermitian point-gap topology in the corresponding non-interacting fermion models after the Jordan-Wigner transformation. Inset: The single-particle spectrum gives nontrivial (trivial) point-gap winding around 0 in the complex plane when the non-Hermitian parameter $\delta$ (the line-gap parameter $\Delta J$) dominates, coinciding with the presence (absence) of nontrivial worldline winding $w_{opt}\neq 0$ ($w_{opt}=0$) under PBCs.}
    \label{DiffDJ}
\end{figure}

\begin{figure}
    \centering
    \includegraphics[width = 1.05\linewidth]{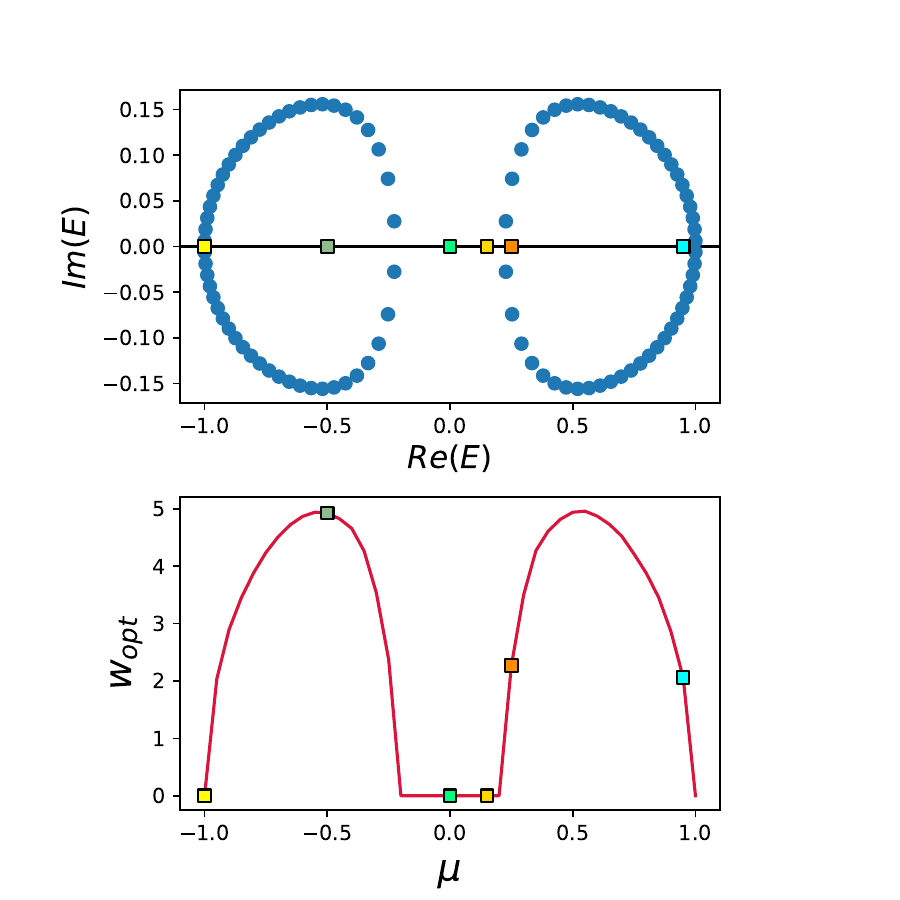}
    \caption{The dominant worldline winding number $w_{opt}$ becomes nonzero when and only when the varying Fermi energy $\mu$, thus the reference energy $E_P$, leads to a nontrivial point-gap topology -  a surrounding loop in the spectrum, accompanied by the closure of the line gap. The colored dots correspond to different reference points $E_P$ for the point-gap topology and Fermi energies $\mu$ of the (Jordan-Wigner transformed) model. We set $\Delta J=0.3$ and $\delta=0.2$.}
    \label{wVersusmu}
\end{figure}

Then, we study the impact of different values of $\Delta J$, and summarize the evolution of the dominant worldline winding number $w_{opt}$, i.e., the Gaussian-fit peak location in the $w$ distribution in Fig. \ref{DiffDJ}. Interestingly, we observe $w_{opt}\neq 0$ if and only if $|\delta| > |\Delta J|$. This parameter space coincides with the nontrivial point-gap topology winding around 0, which guarantees that the non-Hermitian free-fermion chains will display the NHSE under OBCs. On the contrary, when $|\delta| < |\Delta J|$, we have $w_{opt}=0$ despite of nonzero non-Hermitian parameter $\delta$. Here, the SSE-QMC method also needs a boost from enhanced ergodicity for larger $\delta$, especially when $\delta$ surpasses $\Delta J$, consistent with the performances in Fig. \ref{PBCE0_delta}. Such correspondence between nontrivial worldline wining $w_{opt}$ and point-gap topology in the complex spectrum is more apparent in Fig.  \ref{wVersusmu}, as we alter the reference energy $E_P$ by including the term:
\begin{equation}
    H_{\mu}=-\mu \sum_i S_i^z,
\end{equation}
in our model Hamiltonian in Eq. \ref{eq:spinchain}, which corresponds to the Fermi energy $\mu$: $-\mu\sum_i c_i^\dagger c_i$ after the Jordan-Wigner transformation and does not hamper our SSE-QMC algorithm. In full consistency with Fig. \ref{DiffDJ}, whenever the reference energy $E_P=\mu$ falls within a winding spectral loop, we have a nontrivial $w_{opt}\neq 0$. Such a loop also guarantees a vanishing line gap with respect to the reference energy $E_P$ \footnote{The nontrivial point-gap topology must accompany the closure of the line gap with respect to the corresponding reference energy, but not vice versa. For example, for two nested and inverse-flow loops, the spectral winding number with respect to the reference energy inside these loops vanishes, yet the line gap still closes all the time. }. 

\begin{figure}
    \centering
    \includegraphics[width = 1.05\linewidth]{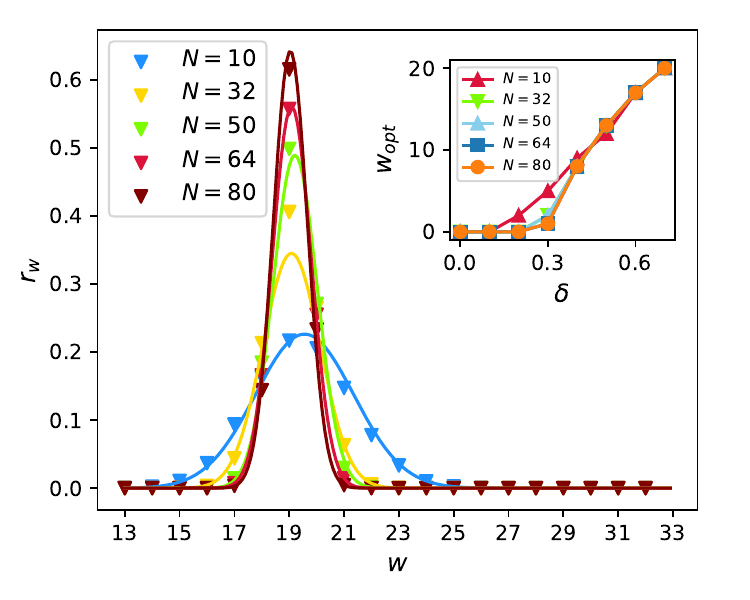}
    \caption{The normalized distribution of worldline winding numbers indicates narrowing Gaussian-fit peaks, while the dominant winding $w_{opt}$ remains almost unchanged, as the system size $N$ increases. Here, $\delta=0.3$ and $\Delta J=0$. Inset: The rise of nontrivial worldline winding at and only at $\delta>\Delta J=0.3$ becomes sharper for larger systems $N$ under PBCs. $\epsilon=0.5$, $J_z=0$ and $\beta=100$. }
    \label{wVersusN}
\end{figure}

It is also interesting to examine the winding-number distributions for various system sizes $N$, which we illustrate in Fig. \ref{wVersusN}. While the width of the distribution relies on $N$, the dominant winding number $w_{opt}$ hardly spots any difference. In large systems, such nontrivial winding consistently introduces global worldlines that traverse the systems and communicate regions far apart, potentially giving rise to long-range quantum entanglement. On the other hand, as we carefully inspect $w_{opt}$ versus $\delta$ for a finite $\Delta J=0.3$, the contrast of zero versus finite $w_{opt}$ across the transition at $\delta_C=\Delta J$ becomes clearer for larger systems, making $w_{opt}$ a better signature for nontrivial point-gap topology as discussed in Fig. \ref{DiffDJ}. 

\begin{figure}
    \centering
    \includegraphics[width = 1.05\linewidth]{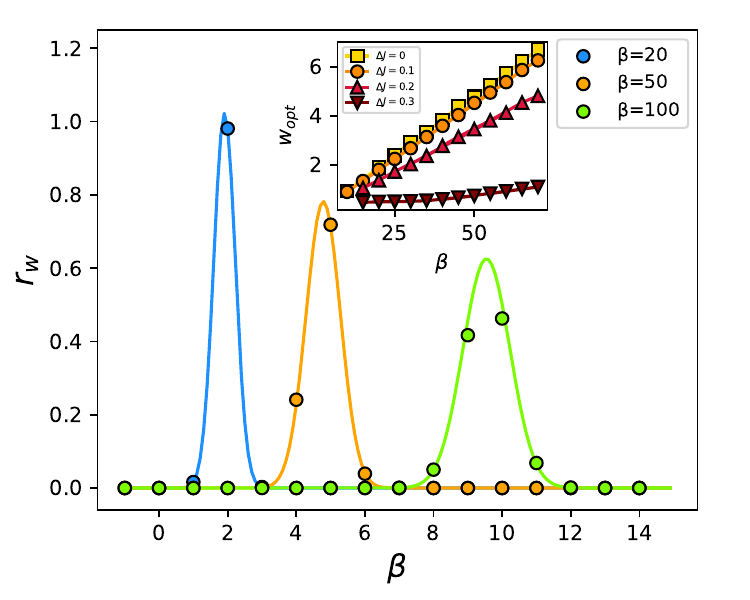}
    \caption{As $\beta$ increases, the distribution of worldline winding numbers for $\Delta J=0$ broadens and shifts towards larger values. The inset shows the Gaussian-fit peak positions $w_{opt}$ versus $\beta$ at different $\Delta J$. In particular, similar to the non-Hermitian toy-model results in Eq. \ref{Partition_w}, $w_{opt}$ depends linearly on $\beta$ for $\Delta J=0$. We fix $\delta = 0.3$, $J_z=0$, $\epsilon=0.5$, and $N=64$ with PBCs. } 
    \label{varbeta}
\end{figure}

Unlike the NHSE, which works only for single-particle eigenstates at zero temperature and without interacting, the nontrivial worldline winding is a quantum phenomenon that straightforwardly generalizes to finite temperatures and interacting systems. For example, we analyze the evolution of worldline winding number distributions in SSE-QMC samples of quantum spin chains for increasing $\beta$. The resulting Gaussian-shaped distributions in Fig. \ref{varbeta} display broadening widths and increasing peak winding number $w_{opt}$ for larger $\beta$. Especially, $w_{opt}$ increases linearly with $\beta$ for models without the line gap $\Delta J$. These features are consistent with the toy-model results in Eq. \ref{Partition_w} and also indicate that, unlike Hermitian quantum systems, we cannot focus solely on the $w=0$ winding-number sector commonly used for SSE-QMC initialization nor equate different winding number sectors in the low-temperature limit $\beta \rightarrow \infty$, as we discussed in Sec. \ref{sec:IIIB}. 

\begin{figure}
    \centering
    \includegraphics[width = 1\linewidth]{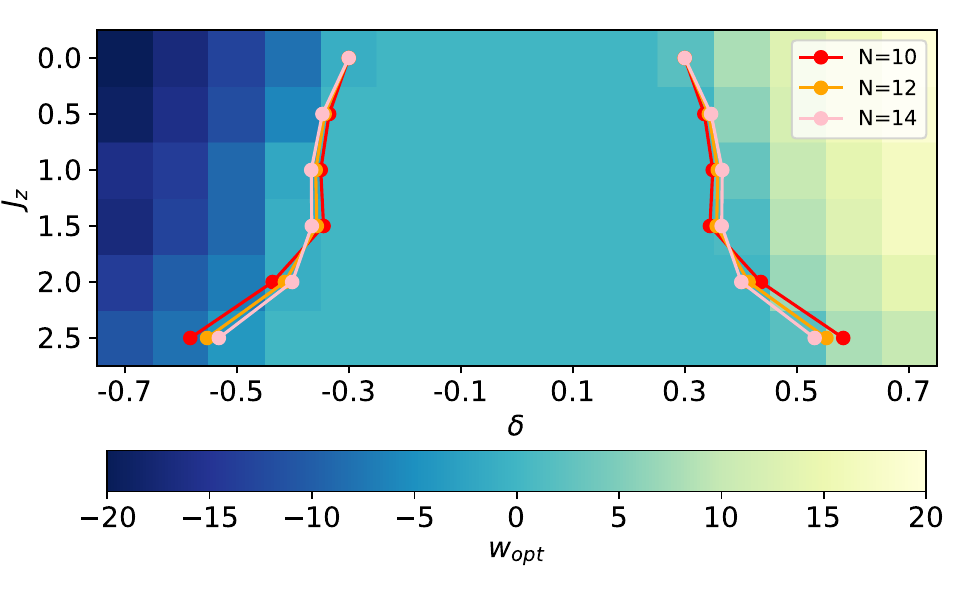}
    \caption{The dominant worldline winding number $w_{opt}$ (color scale) shows explicit dependence on the non-Hermitian parameter $\delta$ as well as the interaction parameter $J_z$, therefore offering characterization of non-Hermitian point-gap topology for interacting quantum systems. We set $\Delta J=0.3$, $\beta=100$, $\epsilon=0.5$, and $N=64$ with PBCs. The dotted lines are the benchmark critical $\delta_C$ from the many-body (single-particle for $J_z=0$) spectrum \cite{manybody10, manybody12} for various $J_z$ and smaller system sizes $N=10, 12, 14$ due to elevated computational cost. }
    \label{diffJz}
\end{figure}

Analysis based upon worldline winding also applies to interacting fermion systems, equivalent to quantum spin chains with nonzero $J_z$ after the Jordan-Wigner transformation in Eq. \ref{eq:jordan}. For instance, we study the worldline winding-number distributions in the SSE-QMC calculations for various $\delta$ and $J_z$ and summarize the dominant $w_{opt}$ in Fig. \ref{diffJz}. In addition to the non-Hermitian parameter $\delta$, the interaction parameter $J_z$ also visibly influences the non-Hermitian topology. The $w_{opt}$ results are also consistent with the (finite-size extrapolations of) many-body spectrum-flow-based identifications \cite{manybody10, manybody12}, also plotted in Fig. \ref{diffJz} as the dotted lines. However, such evaluations commonly require the full spectra and exponential computational costs and thus are applicable only for smaller interacting quantum systems; see Appendix \ref{app:A} for detailed results. We also note that identifying and analyzing such a non-Hermitian quantum many-body system is beyond the NHSE, which requires non-interacting eigenstates under OBCs under the right single-particle eigenstate basis. Consequently, the nontrivial worldline winding offers a novel and efficient characterization of non-Hermitian topology on interacting quantum systems. 


The nontrivial worldline winding number also extends straightforwardly to non-Hermitian quantum systems without translation symmetries. Finally, rather than offering a binary ``yes or no" verdict on the point-gap topology, the finite values of $w_{opt}$, if any, offer a more quantitative measure of the extent of non-Hermitian topology at play. In summary, nontrivial worldline winding offers a broader range of applicability for studying and identifying nontrivial point-gap topology in non-Hermitian quantum systems.

\section{Consequence of nontrivial worldline winding: non-Hermitian entanglement entropy}\label{sec:IV}

\subsection{Entanglement entropy in non-Hermitian quantum systems}\label{sec:IVA}

The nontrivial worldline winding also has immediate physical consequences. For example, such nontrivial winding guarantees worldlines' inevitable passages across boundaries and thus global presence [Fig. \ref{vertexandConfig}(c)], introducing extra entanglement between the regions, even those far apart. We expect these effects to manifest in the real-space entanglement entropy of non-Hermitian quantum systems. In particular, we focus on the Renyi (entanglement) entropy \cite{Calabrese_2009}:
\begin{equation}
    S_{A}^{(n)}=\frac{1}{1-n}\operatorname{ln}\left(\operatorname{Tr} \hat \rho_A^n \right),
\end{equation}
where $\hat \rho_A$ is the (reduced) density matrix of the subsystem $A$. Hereafter, we focus on the second ($n=2$) Renyi entropy $S^{(2)}_A$ \cite{Peschel_2009, IngoPeschel_2003, SSE-Renyi, SSE-Renyi2}. 

However, the definitions of entanglement entropy in non-Hermitian quantum systems remain ambiguous. A simple generalization from the Hermitian case suggests $S^{(2)}_A=-\operatorname{ln}\left( \operatorname{Tr} \hat \rho_A^2 \right)$ \cite{RenyiDefinition}; however, $\hat \rho_A$ is no longer Hermitian, neither is it positive-semidefinite or even real-valued, and the resulting $S^{(2)}_A$ is complex-defined making its meaning as an entanglement measure obscure. On the other hand, the formalism $\Tilde{S}^{(2)}_A=-\operatorname{ln}\left( \operatorname{Tr} |\hat \rho_A|^2 \right)=-\operatorname{ln}\left[ \operatorname{Tr} \left( \hat \rho^\dagger_A \hat \rho_A \right) \right]$ guarantees a positive semi-definite entropy, yet the absolute value is a drastic, non-analytic process. For clarity, we will present results following both definitions, and for each definition, check out the differences $\Delta S$:
\begin{equation}
    \Delta S=S^{(2)}_{A,PBC}-S^{(2)}_{A,OBC},
\end{equation}
between $S^{(2)}_{A,PBC}$ under PBCs and $S^{(2)}_{A,OBC}$ under OBCs with trivial winding $w=0$, and locate the non-Hermitian entanglement entropy contributions accompanying nontrivial worldline winding, and in turn, nontrivial point-gap topology around zero reference energy.

Previously, there have been studies on phase transitions in non-Hermitian quantum systems with entanglement entropy as an indicator: Tu et al. \cite{Chang_NegativeCFT} and Chang et al. \cite{Ryu_nonUnitaryCFT} studied the entanglement entropies under different definitions and revealed the non-unitary conformal field theory with negative central charge $c<0$, as well as the crossover between $c>0$ and $c<0$, in non-Hermitian systems; Chen et al. \cite{Lu_edgeEE} showed the different scaling behavior of von-Neumann entropy at $|\Delta J|>|\delta|$ and $|\Delta J|<|\delta|$ in the non-Hermitian SSH model (Sec. \ref{sec:IIB}), and a negative central charge at the crossover at the critical points $|\Delta J|=|\delta|$; similar concept of edge entanglement entropy $S_{\textrm{edge}}=S_{OBC}-\frac{1}{2}S_{PBC}$ detected the many-body edge states and related phase transitions \cite{Lu_edgeEE, PhysRevB.101.121109}; Guo et al. \cite{Guo_Fermi} discovered the $\operatorname{log}(L)$ scaling of the von-Neumann entanglement entropy with its coefficient related to the Fermi-point topology, which is also consistent with our conclusions. In the following subsection, we will analyze the difference $\Delta S$ between the Renyi entropies under PBC and OBC in direct connection with the nontrivial worldline winding $w_{opt}$ and the corresponding non-Hermitian point-gap topology.

\subsection{Non-Hermitian Entanglement entropy in free-fermion systems}\label{sec:IVB}

Here, we focus on non-Hermitian 1D free-fermion models, equivalent to non-Hermitian quantum spin chains with $J_z=0$. The Hamiltonians take a quadratic form:
\begin{equation}
\hat H= \sum_{i,j} c^{\dagger}_i \mathcal{H}_{ij} c_j = \sum_n \epsilon_{n} |\psi_{n}^R\rangle \langle \psi_{n}^L|, 
\end{equation}
where $c_i$ ($c_i^{\dagger}$) is the fermion annihilation (creation) operator at site $i$, and $|\psi_{n}^R\rangle$ and $\langle \psi_{n}^L|$ are the single-particle biorthogonal basis obtainable from $\mathcal{H}$'s decomposition. For free fermions, we can obtain the single-particle (reduced) density operator $\hat \rho_A$ for region $A$ from the correlation matrix $C_{ij}= \langle c^{\dagger}_i c_j \rangle $ \cite{Peschel_2009, IngoPeschel_2003}, where $i, j\in A$, and subsequently, the second Renyi entropy:
\begin{eqnarray}
    S_A^{(2)}&=&-\sum_{n}\operatorname{ln}\left[\xi^2_n+(1-\xi_n)^2\right], \nonumber \\
    \Tilde S_A^{(2)}&=&-\sum_{n}\operatorname{ln}\left[|\xi_n|^2+(1-|\xi_n|)^2\right],
\end{eqnarray}
where $\xi_n$ are the eigenvalues of the density operator (correlation matrix). To suppress the potential impacts of the edge physics \cite{PhysRevB.101.121109}, we define $A$ as the central region between the $(N/4)^{th}$ site and $(3N/4)^{th}$ site on a chain of length $N$. 

\begin{figure}
    \centering
    \includegraphics[width = 1.15\linewidth]{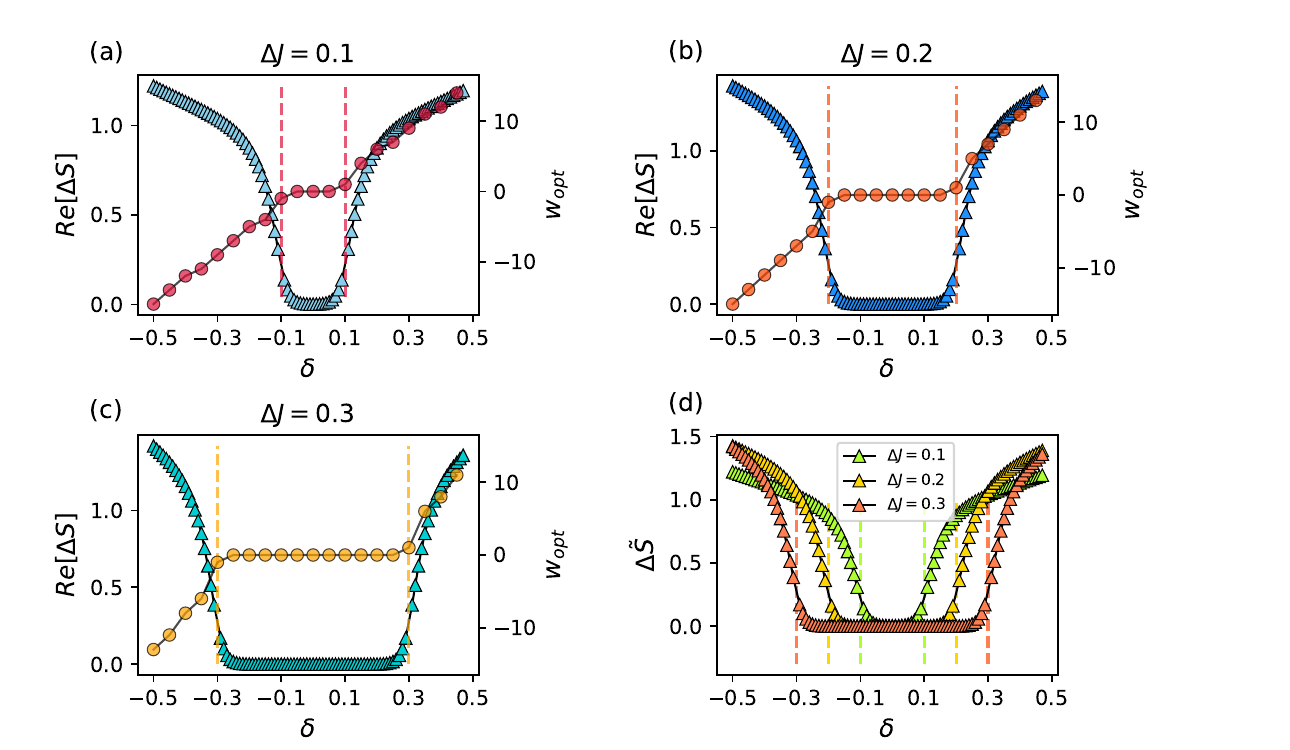}
    \caption{(The real parts of) the differences $\operatorname{Re}\left(\Delta S\right)$ (triangles) between the second Renyi entropy of non-Hermitian free-fermion ground states under PBCs and OBCs exhibit apparent relevance with the dominant winding number $w_{opt}$ (circles), thus the point-gap topology, for (a) $\Delta J=0.1$, (b) $\Delta J=0.2$, and (c) $\Delta J=0.3$, respectively. (d) Following the alternative Renyi entropy definitions $\Tilde{S}^{(2)}_A$, we obtain similar conclusions on $\Delta \Tilde{S}$ between PBCs and OBCs for various $\Delta J$. The dashed lines denote $\delta=\pm \Delta J$ where the transitions locate. $N=64$. }
    \label{Renyi_Free}
\end{figure}

The resulting differences between entanglement entropy under PBCs \footnote{The PBCs for spin chains are equivalent to APBCs for fermion chains after the Jordan-Wigner transformation; see related discussion in Appendix \ref{app:B}.} and OBCs are summarized in Fig. \ref{Renyi_Free}. For various $\Delta J$, we observe consistently vanishing differences $\Delta S$ and $\Delta \Tilde S$ for $-\Delta J<\delta<\Delta J$, where the non-Hermitian quantum systems' worldline winding and the point-gap topology is trivial. Interestingly, positive values of $\Delta S$ and $\Delta \Tilde S$ emerge for $|\delta|>\Delta J$, indicating additional entanglement contributions from nontrivial worldlines with non-zero $w_{opt}$. Drastic changes in $\Delta S$ and $\Delta \Tilde S$ occur in between, which may help locate the topological transitions. Similar studies of non-Hermitian entanglement entropy also apply to quantum systems with interactions and finite temperatures. 

\begin{figure}
    \centering
    \includegraphics[width = 1.05\linewidth]{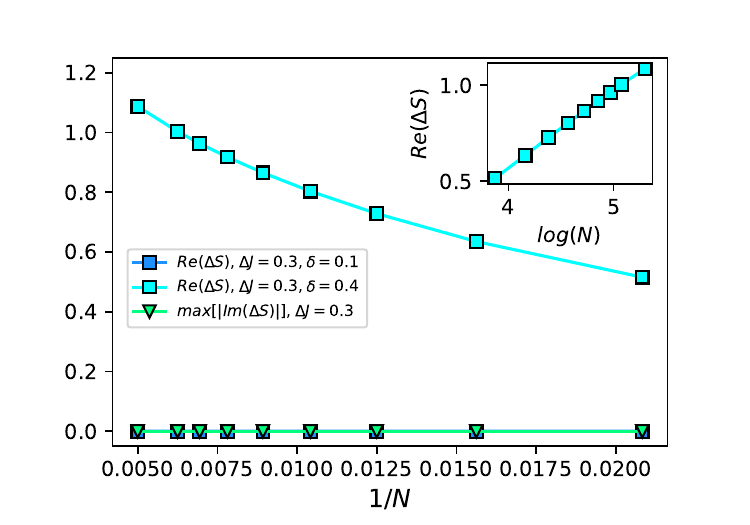}
    \caption{The finite-size scaling of the Renyi entropy difference $\Delta S=S^{(2)}_{A, PBC}-S^{(2)}_{A, OBC}$ suggests that the real part of $\Delta S$ (cyan squares) scales away from zero in the thermodynamic limit $N \rightarrow \infty$ in the presence of nontrivial point-gap topology ($\delta > \Delta J$). In contrast, the real part of $\Delta S$ (blue squares) with trivial point-gap topology ($\delta < \Delta J$) and the imaginary part of $\Delta S$ (green triangles) stays at zero as $N$ increases. For clarity, we show only the largest imaginary-part amplitude of $\Delta S$ among various $\delta$ and $\Delta J=0.3$. $J_z=0$. Inset: The entanglement entropy from nontrivial worldline winding follows a logarithmic scaling: $\Delta S \propto \log(N)$. }
    \label{RenyiScaling}
\end{figure}

We also analyze the finite-size scaling of the Renyi entropy difference $\Delta S$ in Fig. \ref{RenyiScaling}. For non-Hermitian quantum systems without nontrivial worldline winding and point-gap topology, $\Delta S$ tends to zero in the thermodynamic limit as expected. However, in the presence of nontrivial worldline winding, $\Delta S$ possesses a non-zero value and a rising tendency in the $N\rightarrow \infty$ limit, consistent with the results and further asserting the conclusions in Fig. \ref{Renyi_Free}. Notably, such entanglement entropy from nontrivial worldline winding follows a logarithmic scaling with respect to the system size $N$, resembling quasi-long-range entanglement under the Area law with a logarithmic correction in gapless quantum systems \cite{Zhang2011CSL}. While such behavior is qualitatively consistent with our intuition, as the global worldlines persist to large systems (Fig. \ref{wVersusN}) and introduce additional entanglement between regions, even far-apart ones \footnote{At the same time, the changed point-gap topology amounts to the closure of line gaps across the reference point in the complex energy plane, comparable to insulator-metal transitions in Hermitian quantum systems with similar entanglement behaviors. }, more quantitative arguments of the entanglement contribution remains an open question.

\section{Discussion}\label{sec:V}

In summary, we have uncovered the emergent dominance of nontrivial worldline winding in non-Hermitian quantum systems under PBCs, even with interactions, finite temperatures, and various system sizes, which may possess essential impacts on the worldline winding and topology. Empirically, the emergence is in line with, thus offers a broader and more quantitative measure for the non-Hermitian point-gap topology. Unlike the NHSE associated with the right eigenstates, the nontrivial worldline winding exhibits its physical effects as biorthogonal observables, including additional non-Hermitian entanglement entropy. We note that the correspondences between nontrivial worldline winding, point-gap topology, and the potentially quasi-long-range entanglement entropy contributions, though intuitive and reasonable due to their simultaneous global natures, are our hypothesis and established either numerically or based upon toy models. An interesting future direction is to derive more rigorous theoretical connections. 

In the QMC calculations for non-Hermitian quantum systems, such nontrivial worldline winding, together with the barrier between different winding-number sectors, may hamper the ergodicity and proper convergence. For non-Hermitian quantum spin chains, we propose a simple algorithmic remedy to enhance ergodicity between different winding-number sectors. We note that the nontrivial worldline winding is a general phenomenon with clear-cut physical significance and undoubtedly beyond the SSE-QMC formalism, even if we have mainly discussed the worldlines in the SSE-QMC method and used SSE-QMC results for illustrations. Indeed, we have showcased and analyzed the presence of nontrivial worldline winding in the non-Hermitian toy model with the path-integral approach and the non-Hermitian free-fermion models with the exact solutions under the single-particle bases. It will be interesting to investigate analogous worldline winding physics in other QMC and non-QMC algorithms. 

Finally, we have focused on non-Hermitian quantum systems in 1D and the simplest point-gap topology. We note the fascinating possibilities at higher dimensions, with rich categories of non-Hermitian topological phenomena at the research frontier and diverse boundary conditions for worldline windings and braidings. Recently, the NHSE in higher dimensions and its interplay with boundary conditions has attracted much attention \cite{kawabatahigher, okugawa2020, fu2021, yu2021ho, palacios2021, st2022,zhu2022hybrid, zhang2022uni, wang2022amoeba, yokomizo2023nonbloch, james2023unra, hu2023nonhermitian}. However, numerical difficulties in non-Hermitian Hamiltonians, e.g., boundary sensitivity and instability \cite{yang2020, liu2021exact, guo2021exact}, may hamper studies and progress, especially in higher dimensions. Nontrivial worldline winding offers a novel, physically intuitive perspective and better numerical stability under PBCs on such problems. The efficiency and compatibility of the SSE-QMC method in higher dimensions also offer practical research facilities in non-Hermitian quantum systems with interactions and finite temperatures.

\emph{Acknowledgment:} We acknowledge helpful discussions with Lei Wang and Kun Ding. We also acknowledge support from the National Key R\&D Program of China (No.2022YFA1403700) and the National Natural Science Foundation of China (No.12174008 \& No.92270102).

\bibliography{ref.bib}

\appendix


\section{Spectrum-flow-based identification for non-Hermitian topological physics}\label{app:A}

In Sec. \ref{sec:IIID} in the main text, we studied the dominant worldline winding numbers of non-Hermitian quantum systems in the presence of finite interaction $J_z$, such as the results in Fig. \ref{diffJz} in the main text. To establish their connections with the non-Hermitian topology in such interacting systems $\hat{H}$, we evaluate the winding number of the many-body spectrum following Refs. \onlinecite{manybody10, manybody12}:
\begin{equation}
    W(E):=\oint_0^{2 \pi} \frac{d \phi}{2 \pi \mathrm{i}} \frac{d}{d \phi} \log \operatorname{det}\left[\hat{H}(\phi)-E\right], \label{eq:we}
\end{equation}
where $E$ is a reference energy. $\phi \in [0,2\pi)$ denotes the boundary condition and the effective flux through the loop of the system under PBC. 

\begin{figure*}
    \centering
    \includegraphics[width = 1\linewidth]{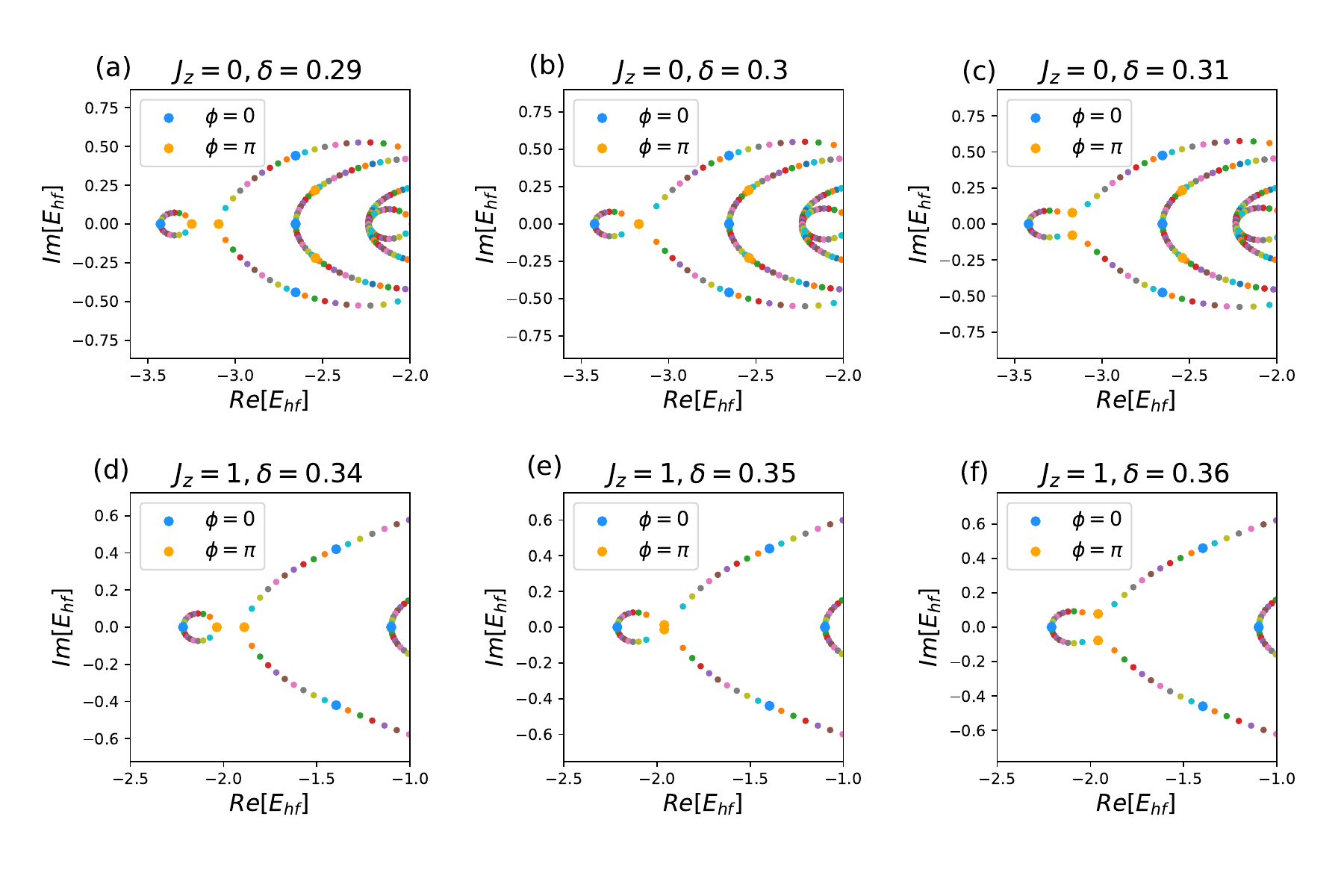}
    \caption{The complex spectra of the overall energy $E_{hf}$ of the non-Hermitian quantum system in Eq. \ref{eq:spinchain} in the main text evolve with the flux $\phi$, as illustrated by various colors. At the critical point of a topological transition, such as (b) $\delta=\Delta J=0.3$, $J_z=0$, and (e) $\delta=0.35$, $\Delta J=0.3$, $J_z=1$, where two $E_{hf}$ touch at $\phi=\pi$, changing the loop topology and the winding number concerning reference energies, say, $E$ at the bottleneck. We have $\Delta J=0.3$ and $N=10$ in all these panels. }
    \label{MBSpectrum}
\end{figure*}

For instance, the complex spectra flow of the overall energy $E_{hf}$ for non-Hermitian quantum systems in Eq. \ref{eq:spinchain} (at half-filling in the fermion representation after the Jordan-Wigner transformation) in the main text is shown in Fig. \ref{MBSpectrum}. As $\phi$ changes and is denoted by different colors, the many-body eigenenergies $E_{hf}$ move in the complex plane. Their trajectories form loops, offering a clear-cut signature for topological transitions as the loops merge or separate and the winding number changes for selected reference energies, e.g., the point where loops touch on the complex $E_{hf}$ plane. Also, such merge or separation of loops in the complex plane of many-body eigenenergies, i.e., whether the ground state and excited states touch or not, is consistent with the absence or presence of line gaps through the reference energy $E_P=0$ in the single-particle picture, which dominates the dominant worldline winding number in the absence of $J_z$.

In particular, for $J_z=0$ that maps to a non-interacting non-Hermitian fermion chain, we locate the transition at $\delta= \Delta J=0.3$ (Fig. \ref{MBSpectrum}a-c), consistent with the single-particle analysis in the presence of a competing line gap $\Delta J$. Further, we generalize the many-body spectrum flows to cases with finite interaction $J_z>0$, e.g., $J_z=0.1$ (Fig. \ref{MBSpectrum}d-f), where the single-particle analysis no longer applies. We also observe that the separate loops at smaller $|\delta|$ merge together at larger $|\delta|$, yet the transition where the loops touch occurs at $|\delta|>|\Delta J|$ with finite interaction $J_z>0$. We note that the evaluation of Eq. \ref{eq:we} requires repeated matrix operations of many-body Hamiltonians, whose cost increases exponentially with the system size, limiting its applicability to small systems. 

For $N=10, 12, 14$, fixed $\Delta J=0.3$, and various values of $J_z$, we track the many-body spectra flows and determine the phase boundaries in $\delta$, as the dotted lines in Fig. \ref{diffJz} in the main text. The results are consistent with the appearance of the nontrivial SSE-QMC worldline winding we have described in the main text.

\section{ Boundary condition and entanglement entropy of corresponding spin and fermion chains}\label{app:B}

While we employ PBC for the spin chain in the main text, the boundary condition of the corresponding fermion chain following the Jordan-Wigner transformation needs to be treated carefully \cite{Nagaosa1999QuantumFT}:
\begin{eqnarray}
    S_N^+S_1^-&=&-Kf^{\dagger}_Nf_1, \\\nonumber
    K&=&\operatorname{exp}\left[i\pi\sum_{i=1}^N f^{\dagger}_if_i\right]=(-1)^{N_f},
\end{eqnarray}
where $N_f$ is the total number of fermions. As a result, for a spin chain with $S_{tot}^z=0$ and PBC, the boundary condition of the corresponding fermion chain at half filling is anti-periodic for $N=4m$ and periodic for $N=4m+2$, $m\in\mathbb{Z}$. Alternatively, for a fermion chain with PBC, the corresponding spin chain should obey the anti-periodic boundary condition (APBC) for $N=4m$. 

\begin{figure}
    \centering
    \includegraphics[width = 1.15\linewidth]{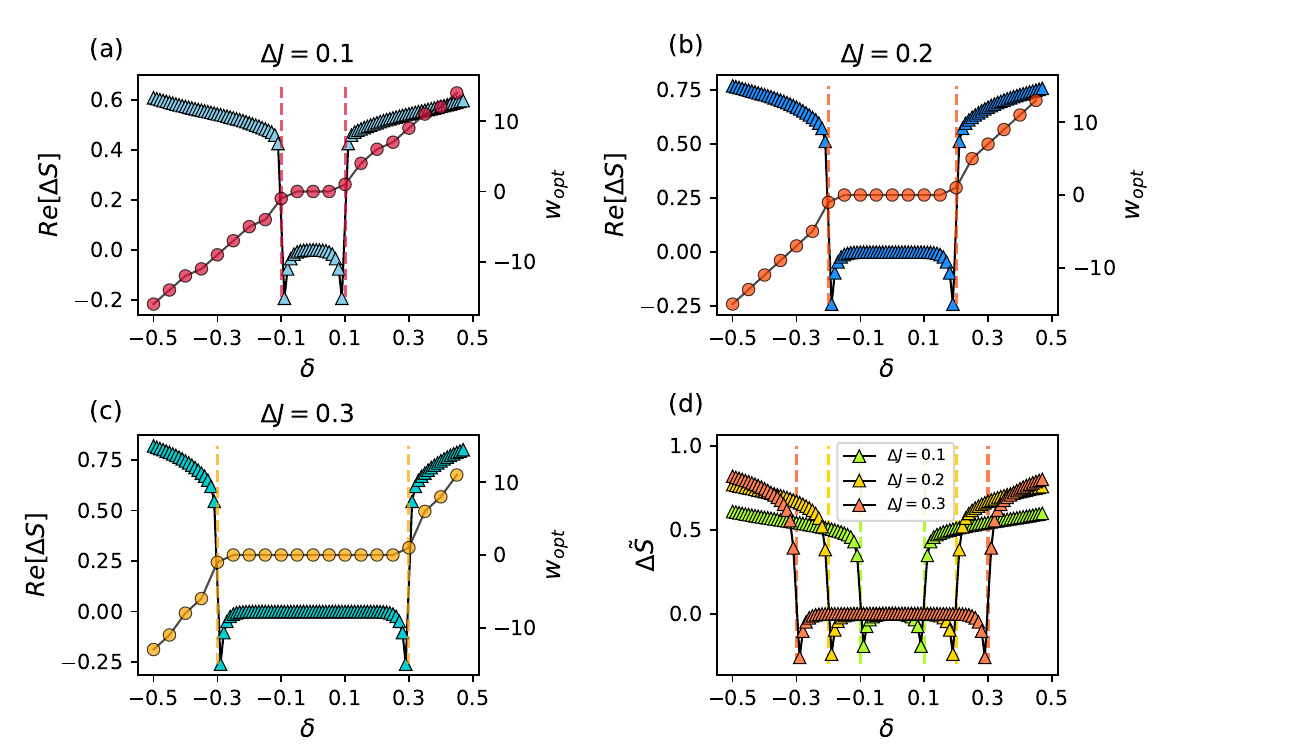}
    \caption{(The real parts of) the differences $\operatorname{Re}\left(\Delta S\right)$ (triangles) between the second Renyi entropy of non-Hermitian free-fermion ground states under APBCs and OBCs exhibit apparent relevance with the dominant winding number $w_{opt}$ (circles), thus the point-gap topology, for (a) $\Delta J=0.1$, (b) $\Delta J=0.2$, and (c) $\Delta J=0.3$, respectively. (d) Following the alternative Renyi entropy definitions $\Tilde{S}^{(2)}_A$, we obtain similar conclusions on $\Delta \Tilde{S}$ between APBCs and OBCs for various $\Delta J$. The dashed lines denote $\delta=\pm \Delta J$ where the transitions locate. $N=64$. } 
    \label{Renyi_Free_APBC}
\end{figure}

\begin{figure}
    \centering
    \includegraphics[width = 1.05\linewidth]{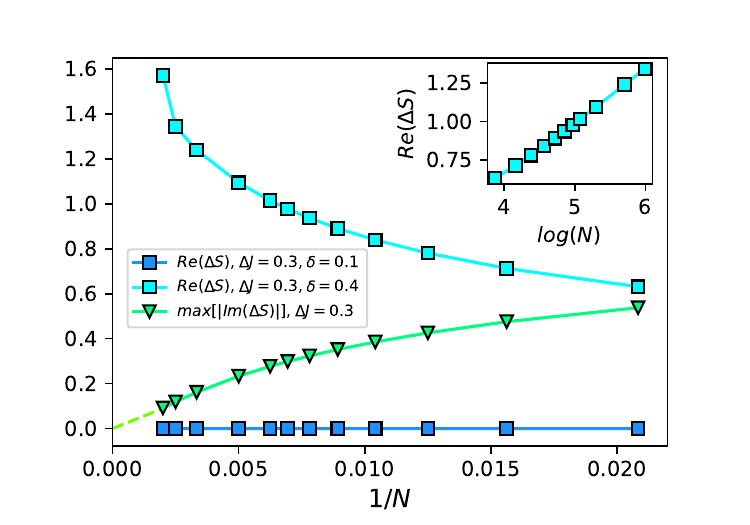}
    \caption{The finite-size scaling of the Renyi entropy difference $\Delta S=S^{(2)}_{A, APBC}-S^{(2)}_{A, OBC}$ suggests that the real part of $\Delta S$ (cyan squares) scales away from zero in the thermodynamic limit $N \rightarrow \infty$ in the presence of nontrivial point-gap topology ($\delta > \Delta J$). In contrast, the real part of $\Delta S$ (blue squares) with trivial point-gap topology ($\delta < \Delta J$) and the imaginary part of $\Delta S$ (green triangles) generally scale to zero. For clarity, we show only the largest imaginary-part amplitude of $\Delta S$ among various $\delta$ and $\Delta J=0.3$. $J_z=0$. Inset: The entanglement entropy from nontrivial worldline winding follows a logarithmic scaling: $\Delta S \propto \log(N)$. } 
    \label{Renyi_scaling_withflux}
\end{figure}

In the main text, we have shown the results on the differences $\Delta S$ of the $2^{nd}$ Renyi entropy between PBC and OBC spin chains (i.e., APBC and OBC fermion chains) at $N=64$. Here, we show in Fig. \ref{Renyi_Free_APBC} the difference between APBC and OBC spin chains (i.e., PBC and OBC fermion chains) at the same system size. Despite some differences at the phase transitions $|\Delta J|=|\delta|$, the overall behavior of the entanglement entropy is consistent with the trivial or nontrivial worldline winding for both PBC and APBC. Similarly, we show the finite-size scaling of the difference of Renyi entropy $\Delta S$ between APBC and OBC spin chains in Fig. \ref{Renyi_scaling_withflux}. We note the quasi-long-range entanglement corresponding to the logarithmic entanglement-entropy scaling in the presence of nontrivial point-gap topology, the vanishing line gap, and the nontrivial worldline winding appears irrespective of whether we consider PBC or APBC; however, the imaginary part of $\Delta S$ scales to zero under APBC, rather than strictly being zero under PBC.

\end{document}